%% file: idtdr1.tex
\documentclass[longauth]{aa} % for the long lists of affiliations 
\usepackage{amsmath}
\usepackage{graphicx}
\usepackage{txfonts}
\usepackage{macros-idtdr1}

\begin{document}

   \title{Gaia Data Release 1}

   \subtitle{Pre-processing and source list creation}

%  \input{authors-trial}
   \input{authors}

   \date{ }

% \abstract{}{}{}{}{} 
% 5 {} token are mandatory
 
% \abstract

\abstract{
The first data release from the Gaia mission contains
accurate positions and magnitudes for more than a billion sources, and 
proper motions and parallaxes for the majority of the 2.5~million Hipparcos and Tycho-2
stars.
}{
We describe three essential elements of the initial data treatment leading to
this catalogue: the image analysis, the construction of a source list,
and the near real-time monitoring of the payload health. We also discuss some
weak points that set limitations for the attainable precision at the present
stage of the mission.
}{
Image parameters for point sources are derived from one-dimensional scans,
using a maximum likelihood method, under the assumption of a line spread
function constant in time, and a complete modelling of bias and background.
These conditions are, however, not completely fulfilled.
The Gaia source list is built starting from a large ground-based catalogue,
but even so a significant number of new entries have been added, and a large
number have been removed. The autonomous onboard star image detection will pick up
many spurious images, especially around bright sources, and such unwanted
detections must be identified. Another key step of the source list creation
consists in arranging the more than $10^{10}$ individual detections in
spatially isolated groups that can be analysed individually.
}{
Complete software systems have been built for the Gaia initial data treatment,
that manage approximately 50~million focal plane transits daily, giving transit
times and fluxes for 500~million individual CCD images to the astrometric and 
photometric processing chains. 
The software also carries out a successful and detailed daily monitoring
of Gaia health.  
}{} 


   \maketitle
%
%________________________________________________________________

\input{introduction}

\input{attitude}

\input{ipd_intro}

\input{ipd_instrument}

\input{ipd_sso}
\input{ipd_xp}
\input{ipd_mll}

\input{xm_intro}
\input{xm_igsl}
\input{xm_scene}

\input{xm_spurious}

\input{xm_coord}

\input{xm_mcg}

\input{xm_solve}

\input{bam}

\input{valid_mon}

\input{valid_fl}

\input{valid_avubam}

\input{valid_avuaim}

%
%______________________________________________________________

\section{Conclusions}

Software systems have been built for the pre-processing of the Gaia
observations, for fundamental scientific payload calibrations, and for
monitoring the payload health and data quality, including independent
verifications. Together with a detection classification and a global
cross-match process, these systems form the bases for the downstream
astrometric and photometric processing chains leading to Gaia-DR1.

At this early stage of the mission, the data processing is still not
fully deployed. Of special relevance for Gaia-DR1 is that the fainter
component of close double stars is not processed, the instrument model
does not consider chromatic effects, and the cross match has not caught
stars with very high proper motion. Also CTI effects are not considered,
and in combination with chromaticity, this may lead to minor biases in
sky regions where scan directions are not well distributed.

\begin{acknowledgements}

This work has made use of data from the ESA space mission Gaia, 
processed by the Gaia Data Processing and Analysis Consortium
(DPAC). Funding for the DPAC has been provided by national institutions, in
particular the institutions participating in the Gaia Multilateral Agreement.
The Gaia mission website is: \url{http://www.cosmos.esa.int/gaia}.

The authors are members of the DPAC,
and this work has been supported by the following funding agencies: 

MINECO (Spanish Ministry of Economy) - FEDER through grant ESP2013-48318-C2-1-R
and ESP2014-55996-C2-1-R and MDM-2014-0369 of ICCUB (Unidad de Excelencia
`Mar\'ia de Maeztu'); 

The Netherlands Research School for Astronomy (NOVA); 

the German Aerospace Agency DLR under grants 50QG0501, 50QG1401, 50QG0601, 50QG0901, and 50QG1402; 

the European Space Agency in the framework of the Gaia project; 

the Agenzia Spaziale Italiana (ASI) through grants  ASI  I/037/08/0, ASI I/058/10/0, ASI 2014-025-R.0, and ASI 2014-025-R.1.2015 and the Istituto Nazionale di AstroFisica (INAF); 

the United Kingdom Space Agency; 

Funda\c{c}\~ao para a Ci\^{e}ncia e a Tecnologia through the contract Ci\^{e}ncia2007 and project grant PTDC/CTE-SPA/118692/2010.

The authors thankfully acknowledge the computer resources from MareNostrum,
technical expertise and assistance provided by the Red Espa\~nola de
Supercomputaci\'on at the Barcelona Supercomputing Center, Centro Nacional de
Supercomputaci\'on. We also acknowledge the computer resources and support
of CSUC, Consorci de Serveis Universitaris de Catalunya.

In addition to the authors of this work there are many other people who have made valuable contributions to the core processing 
but who have meanwhile moved on to other projects. Among these, we want to specifically mention 
Sebastian Els, 
John Hoar,
Ralf Keil, 
Ralf Kohley,
Bel\'en L\'opez Mart\'i, 
Alexandros Ouzounis, 
Davide Padeletti, 
Michael Perryman, 
Stefan Theil, 
Pau Vall\`es, 
and 
Floor van Leeuwen.  

We also wish to thank
Jos de Bruijne for support since the early developments
and 
Klaus Meisenheimer for organising observing time at Calar Alto for the Gaia Ecliptic Poles Catalogue.
Finally, we thank the referee, Dr Norbert Zacharias, for many constructive
comments and suggestions to an earlier version of this paper.
\end{acknowledgements}

\bibliographystyle{aa} % style aa.bst
\bibliography{refs} % your references refs.bib

\vfill
\input{appendix}

\end{document}

%% file: authors.tex
% Reference person for this section: BAS

% 104 names as of March 10, 12:03 !
% 107 names as of July   7, 17:43 !

\author{C.~Fabricius\inst{\ref{inst:ieecfabri}} % as co-ordinator and main writer, important contributor to many aspects 
\and U.~Bastian\inst{\ref{inst:ari}} % as writer and CU3 leader
\and J.~Portell\inst{\ref{inst:ieec}} % as writer and actual leader of IDT
\and J.~Casta\~neda\inst{\ref{inst:ieec}} % as writer and actual leader of XM
\and M.~Davidson\inst{\ref{inst:uoe}} % as writer and major calibs contributor
\and N.C.~Hambly\inst{\ref{inst:uoe}} % as writer and major calibs contributor
\and M.~Clotet\inst{\ref{inst:ieec}} % as writer and XM contributor
\and M.~Biermann\inst{\ref{inst:ari}} % as writer and FL leader
\and A.~Mora\inst{\ref{inst:auroraesac}} % as writer for BAM
\and D.~Busonero\inst{\ref{inst:oato}} % as writer and AIM leader
\and A.~Riva\inst{\ref{inst:oato}} % as writer and AVU/BAM leader
\and A.G.A.~Brown\inst{\ref{inst:leiden}} % as writer and photometry contributor
\and R.~Smart\inst{\ref{inst:oato}} % as writer and for the IGSL
\and U. Lammers\inst{\ref{inst:esac}} % as CU3-T and SOC manager
\and J.~Torra\inst{\ref{inst:ieec}} % as manager of IDT/IDU, responsible for the Spanish contribution, and as CU3 deputy
%\and M.G.~Lattanzi\inst{\ref{inst:oato}} % as AVU manager, co-writer on AIM and AVU-BAM, responsible for ASI of the DPCT, and as CU3 deputy
\and R.~Drimmel\inst{\ref{inst:oato}} % as AVU leader, co-writer on AIM and AVU-BAM
\and G.~Gracia\inst{\ref{inst:poesac}} % for his role in the SCI management
%\and NOT S. Els - he said it is not necessary.
\and W.~L{\"o}ffler\inst{\ref{inst:ari}} % as FL technical leader and architect
\and A.~Spagna\inst{\ref{inst:oato}} % as XM contributor and former leader
\and L.~Lindegren\inst{\ref{inst:lund}} % for many major contributions, Attitude, LSF, GT, MLL, ... 
\and S.~Klioner\inst{\ref{inst:tud}} % LATT
%
%\and the rest of the IDT/IDU/FL/DPCE/DPCB/DPCT core crews and contributors in alphabetic sequence:
\and A.~Andrei\inst{\ref{inst:rio}} % contribution to IGSL/QSO
\and N.~Bach\inst{\ref{inst:auroraesac}} % IDT/FL operator
\and L.~Bramante\inst{\ref{inst:altec}} %DPCT operations lead
\and T.~Br{\"u}semeister\inst{\ref{inst:ari}} %FL
\and G.~Busso\inst{\ref{inst:ioa},\ref{inst:leiden}} % as photometry contributor
\and J.M.~Carrasco\inst{\ref{inst:ieec}} % colour-clour relations used in IPD
\and M.~Gai\inst{\ref{inst:oato}} % AVU
\and N.~Garralda\inst{\ref{inst:ieec}}  %IDT/IDU 
\and J.J.~Gonz\'alez-Vidal\inst{\ref{inst:ieec}}  %IDT/IDU 
\and R.~Guerra\inst{\ref{inst:esac}} % DPCE leader
\and M.~Hauser\inst{\ref{inst:ari}} %FL
\and S.~Jordan\inst{\ref{inst:ari}} %FL
\and C.~Jordi\inst{\ref{inst:ieec}} % colour-clour relations used in IPD 
\and H.~Lenhardt\inst{\ref{inst:ari}} %FL
%\and H.~Manche\inst{\ref{inst:imcce}} %No! explicit no in email from Bouquillon of 25.3. %for the SSO contributions
\and F.~Mignard\inst{\ref{inst:oca}} % for the SSO contributions, e.g. the AlphaDelta/Motions validations)
\and R.~Messineo\inst{\ref{inst:altec}} %DPCT manager
\and A.~Mulone\inst{\ref{inst:altec}} %DPCT DB manager
\and I.~Serraller\inst{\ref{inst:ieec}, \ref{inst:gmvesac}} % former IDT lead developer
\and U.~Stampa\inst{\ref{inst:ari}} %FL
\and P.~Tanga\inst{\ref{inst:oca}} % for the SSO contributions, e.g. the AlphaDelta/Motions validations)
\and A.~van~Elteren\inst{\ref{inst:leiden}} % as photometry contributor
\and W.~van Reeven\inst{\ref{inst:auroraesac}} % current IDT lead developer
\and H.~Voss\inst{\ref{inst:ieec}} % for his contributions to the saturation thresholds for the individual CCDs
%
%\and the rest of CU3 members (incl a few former ones) in alphabetic sequence:
\and U.~Abbas\inst{\ref{inst:oato}} 
\and W.~Allasia\inst{\ref{inst:eurix}}
\and M.~Altmann\inst{\ref{inst:ari},\ref{inst:paris}} % yes, two!
\and S.~Anton\inst{\ref{inst:porto},\ref{inst:lisbon}}
\and C.~Barache\inst{\ref{inst:paris}}
\and U.~Becciani\inst{\ref{inst:catania}} % AVU
\and J.~Berthier\inst{\ref{inst:imcce}}
\and L.~Bianchi\inst{\ref{inst:eurix}}
\and A.~Bombrun\inst{\ref{inst:hespaceesac}}
\and S.~Bouquillon\inst{\ref{inst:paris}}
\and G.~Bourda\inst{\ref{inst:bordeaux}}
\and B.~Bucciarelli\inst{\ref{inst:oato}}
\and A.~Butkevich\inst{\ref{inst:tud}}
\and R.~Buzzi\inst{\ref{inst:oato}}
\and R.~Cancelliere\inst{\ref{inst:utorino}}
\and T.~Carlucci\inst{\ref{inst:paris}}
\and P.~Charlot\inst{\ref{inst:bordeaux}}
\and R.~Collins\inst{\ref{inst:uoe}}
\and G.~Comoretto\inst{\ref{inst:vegaesac}}
%\and L.~Corcione\inst{\ref{inst:oato}}
\and N.~Cross\inst{\ref{inst:uoe}}
\and M.~Crosta\inst{\ref{inst:oato}}
\and F.~de~Felice\inst{\ref{inst:padova}}
\and A.~Fienga\inst{\ref{inst:oca}}
\and F.~Figueras\inst{\ref{inst:ieec}}
\and E.~Fraile\inst{\ref{inst:rheaesac}}
\and R.~Geyer\inst{\ref{inst:tud}}
\and J.~Hernandez\inst{\ref{inst:esac}}
\and D.~Hobbs\inst{\ref{inst:lund}}
\and W.~Hofmann\inst{\ref{inst:ari}}
\and S.~Liao\inst{\ref{inst:oato},\ref{inst:shanghai}}
\and E.~Licata\inst{\ref{inst:eurix}}
%\and D.~Loreggia\inst{\ref{inst:eurix}}
\and M.~Martino\inst{\ref{inst:altec}}
\and P.J.~McMillan\inst{\ref{inst:lund}}
\and D.~Michalik\inst{\ref{inst:lund}}
\and R.~Morbidelli\inst{\ref{inst:oato}}
\and P.~Parsons\inst{\ref{inst:theserverlabsesac}}
\and M.~Pecoraro\inst{\ref{inst:eurix}}
\and M.~Ramos-Lerate\inst{\ref{inst:vitrocisetesac}}
\and M.~Sarasso\inst{\ref{inst:oato}}
\and H.~Siddiqui\inst{\ref{inst:vegaesac}}
\and I.~Steele\inst{\ref{inst:liverpool}}
\and H.~Steidelm\"{u}ller\inst{\ref{inst:tud}}
\and F.~Taris\inst{\ref{inst:paris}}
\and A.~Vecchiato\inst{\ref{inst:oato}}
%
%\and some non-CU3-members with heavy contrib to daily pipeline:
\and A.~Abreu\inst{\ref{inst:vegaesac}} % non-CU3 FL Scientist
\and E.~Anglada\inst{\ref{inst:sercoesac}} %non-CU3 member with heavy contrib to daily pipeline
\and S.~Boudreault\inst{\ref{inst:mssl},\ref{inst:mpg}} % non-CU3, PEM-NU 
\and M.~Cropper\inst{\ref{inst:mssl}} % non-CU3, PEM-NU and other CCD aspects
\and B.~Holl\inst{\ref{inst:geneva}} %ex-CU3 member; heavy contribution to AGISLab etc.
\and N.~Cheek\inst{\ref{inst:sercoesac}} %non-CU3 member with heavy contrib to daily pipeline
\and C.~Crowley\inst{\ref{inst:he-spaceesac}} % non-CU3 FL Scientist
\and J.M.~Fleitas\inst{\ref{inst:auroraesac}} % non-CU3 FL Scientist
\and A.~Hutton\inst{\ref{inst:auroraesac}} %non-CU3 member with heavy contrib to daily pipeline
\and J.~Osinde\inst{\ref{inst:isdefeesac}} %non-CU3 member with heavy contrib to daily pipeline
\and N.~Rowell\inst{\ref{inst:uoe}} % PSF contributions, some figures etc.
\and E.~Salguero\inst{\ref{inst:isdefeesac}} %non-CU3 member with heavy contrib to daily pipeline
\and E.~Utrilla\inst{\ref{inst:auroraesac}} %non-CU3 member with heavy contrib to daily pipeline
\and N.~Blagorodnova\inst{\ref{inst:ieec},\ref{inst:caltech}} % significant IDT/IDU contributions
\and M.~Soffel\inst{\ref{inst:tud}} %REMAT and AGIS
\and J.~Osorio\inst{\ref{inst:porto}} % pt responsible
\and D.~Vicente\inst{\ref{inst:bsc}} % DPCB, BSC
\and J.~Cambras\inst{\ref{inst:csuc}} %DPCB, CSUC
\and H.-H.~Bernstein$^\dagger$\inst{,\ref{inst:ari}}
}

%Finally, here is the list of active CU3 members to be excluded from authorship lists: (no DPAC/CU3 contrib, but members for other/technical reasons)
%Rajesh Kumar
%Piero Ranalli
%L.Balaguer-Nuñez
%S.Girona
%A.Gurpide
%S.Rago
%C.Stephenson
%G.Thimm
%N.Kudryavtseva
%A.Schmidt\inst{\ref{inst:esac}}

\institute{
Institut de Ci\`encies del Cosmos,
Universitat de Barcelona (IEEC-UB), Mart\'i Franqu\`es 1, E-08028 Barcelona, 
Spain\ \email{claus.fabricius@am.ub.es}\label{inst:ieecfabri}
\and
Astronomisches Rechen-Institut, Zentrum f\"ur Astronomie der Universit\"at Heidelberg, M\"onchhofstra{\ss}e 14, D-69120 Heidelberg,
Germany
%\ \email{bastian@ari.uni-heidelberg.de}
\label{inst:ari}
\and
Institut de Ci\`encies del Cosmos,
Universitat de Barcelona (IEEC-UB), Mart\'i Franqu\`es 1, E-08028 Barcelona, 
Spain\ \label{inst:ieec}
\and
Institute for Astronomy, School of Physics and Astronomy, University of Edinburgh, 
Royal Observatory, Blackford Hill, Edinburgh, EH9~3HJ, 
United Kingdom
%\ \email{nch@roe.ac.uk;mdv@roe.ac.uk}
\label{inst:uoe}
\and
Aurora Technology for ESA/ESAC, Camino Bajo del Castillo s/n, 28691 Villanueva de la Ca{\~n}ada, Spain
\label{inst:auroraesac}
\and
Istituto Nazionale di Astrofisica, Osservatorio Astrofisico di Torino, Via Osservatorio 20, Pino Torinese, Torino, 10025, 
Italy
%\ \email{busonero@oato.inaf.it;riva@oato.inaf.it;\\lattanzi@oato.inaf.it;drimmel@oato.inaf.it}
\label{inst:oato}
\and
Sterrewacht Leiden, Leiden University, P.O.\ Box 9513, 2300 RA, Leiden, The Netherlands
%\ \email{brown@strw.leidenuniv.nl; vanelteren@strw.leidenuniv.nl}
\label{inst:leiden}
%\and
%GMV, Isaac Newton, 11 PTM, Tres Cantos 28760 Madrid, Spain\label{inst:gmv}
% \e-mail{iserraller@gmv.com}
\and
ESA, European Space Astronomy Centre, Camino Bajo del Castillo s/n, 28691 Villanueva de la Ca{\~n}ada, Spain\label{inst:esac}
\and
Gaia Project Office for DPAC/ESA, Camino Bajo del Castillo s/n, 28691 Villanueva de la Ca{\~n}ada, Spain
\label{inst:poesac}
\and
Lund Observatory, Department of Astronomy and Theoretical Physics, Lund University, Box 43, 22100, Lund, Sweden
\label{inst:lund}
\and
Technische Universit\"{a}t Dresden, Mommsenstrasse 13, D-01062 Dresden,
Germany
\label{inst:tud}
\and
GEA-Observatorio National/MCT,Rua Gal. Jose Cristino 77, CEP 20921-400,
Rio de Janeiro, Brazil
\label{inst:rio}
\and
Altec, Corso Marche 79, Torino, 10146 Italy
\label{inst:altec}
\and
Institute of Astronomy, University of Cambridge, Madingley Road, Cambridge CB3OHA, United Kingdom
%\ \email{giorgia@ast.cam.ac.uk}
\label{inst:ioa}
\and
Observatoire de la C\^{o}te d'Azur,	BP 4229, Nice Cedex 4, 06304, France
\label{inst:oca}
\and
GMV for ESA/ESAC, Camino Bajo del Castillo s/n, 28691 Villanueva de la Ca{\~n}ada, Spain
\label{inst:gmvesac}
\and
EURIX S.r.l., via Carcano 26, Torino, 10153, Italy
\label{inst:eurix}
\and
SYRTE, Observatoire de Paris, PSL Research University, CNRS, Sorbonne Universités, UPMC Univ. Paris 06, LNE, 61 avenue de l’Observatoire, 75014 Paris, France
\label{inst:paris}
\and 
Universidade do Porto, Rua do Campo Alegre, 687, Porto, 4169-007, Portugal
\label{inst:porto}
\and
Institute of Astrophysics and Space Sciences
Faculdade de Ciencias, Campo Grande, PT1749-016 Lisboa, Portugal
\label{inst:lisbon}
\and
INAF, Osservatorio Astrofisico di Catania, Catania, Italy
\label{inst:catania}
\and
IMCCE, Institut de Mecanique Celeste et de Calcul des Ephemerides,77 avenue Denfert-Rochereau, 75014 Paris, France
\label{inst:imcce}
\and
HE Space Operations BV for ESA/ESAC, Camino Bajo del Castillo s/n, 28691 Villanueva de la Ca{\~n}ada, Spain
\label{inst:hespaceesac}
\and
Univ. Bordeaux, LAB, UMR 5804, F-33270, Floirac, France, 
and CNRS, LAB, UMR 5804, F-33270, Floirac, France
\label{inst:bordeaux}
\and
Dipartimento di Informatica, Università di Torino, C.so Svizzera 185, 10149 Torino, Italy
\label{inst:utorino}
\and
Telespazio Vega UK Ltd for ESA/ESAC, Camino Bajo del Castillo s/n, 28691 Villanueva de la Ca{\~n}ada, Spain
\label{inst:vegaesac}
\and
University of Padova, Via Marzolo 8, Padova, I-35131, Italy
\label{inst:padova}
\and
RHEA for ESA/ESAC, Camino Bajo del Castillo s/n, 28691 Villanueva de la Ca{\~n}ada, Spain
\label{inst:rheaesac}
\and
Shanghai Astronomical Observatory, Chinese Academy of Sciences, 80 Nandan Rd, 200030 Shanghai, China
\label{inst:shanghai}
\and
The Server Labs S.L. for ESA/ESAC, Camino Bajo del Castillo s/n, 28691 Villanueva de la Ca{\~n}ada, Spain
\label{inst:theserverlabsesac}
\and
Vitrociset Belgium for ESA/ESAC, Camino Bajo del Castillo s/n, 28691 Villanueva de la Ca{\~n}ada, Spain
\label{inst:vitrocisetesac}
\and
Astrophysics Research Institute, ic2 - Liverpool Science Park, 146 Brownlow Hill, Liverpool L3 5RF, United Kingdom
\label{inst:liverpool}
\and
Serco Gestion de Negocios S.L. for ESA/ESAC, Camino Bajo del Castillo s/n, 28691 Villanueva de la Ca{\~n}ada, Spain
\label{inst:sercoesac}
\and
Mullard Space Science Laboratory, University College London, Holmbury St Mary, Dorking, Surrey RH5\,6NT, United Kingdom
\label{inst:mssl}
\and
Max Planck Institute for Solar System Research, Justus-von-Liebig-Weg 3, 37077 G\"ottingen, Germany
\label{inst:mpg}
\and
Observatoire de Geneve,	chemin des Maillettes 51, Sauverny, CH-1290, Switzerland
\label{inst:geneva}
%\and
%University of Leicester, University Road, Leicester, LE1 7RH, United Kingdom
%\label{inst:leicester}
%\and
%Astronomical Institute, Bern University, Sidlerstrasse 5, Bern, CH-3012, Switzerland
%\label{inst:bern}
\and
HE Space Operations BV for ESA/ESAC, Camino Bajo del Castillo s/n, 28691 Villanueva de la Ca{\~n}ada, Spain
\label{inst:he-spaceesac}
\and
ISDEFE for ESA/ESAC, Camino Bajo del Castillo s/n, 28691 Villanueva de la Ca{\~n}ada, Spain
\label{inst:isdefeesac}
\and
Cahill Center for Astrophysics
California Institute of Technology, Pasadena, CA 91125, USA
\label{inst:caltech}
\and
Barcelona Supercomputing Center, Nexus II Building, Jordi Girona 29, 08034 Barcelona, Spain
\label{inst:bsc}
\and
Consorci de Serveis Científics i Academics de Catalunya (CSUC),  Gran Capita 2, 08034 Barcelona, Spain 
\label{inst:csuc}
}

%% file: introduction.tex
\section{Introduction}

% Reference person for this section: CF

The European Space Agency (ESA) mission Gaia (Gaia collaboration, Prusti et al.\  2016, this volume)
is producing a three-dimensional map of a representative sample of our Galaxy, 
that contains
detailed astrometric and photometric information for more than one billion
stars and solar system objects, as well as galaxies and quasars. It started
nominal operations in July 2014, and a data release, Gaia-DR1, based on the
first 14 months of observations, is now being published (Gaia collaboration,
Brown et al.\ 2016, this volume).

In the present paper we describe the main steps of the initial data treatment,
while the astrometric and photometric reductions are presented elsewhere
(Lindegren et al.\ 2016, this volume; van Leeuwen et al.\ 2016, this volume). 
We give special emphasis to the limitations of the present processing, which will
have an impact on the released data. In future releases, these shortcomings
will gradually disappear.

\section{The Gaia instrument\label{sec:inst}}

The Gaia instrument consists of two telescopes with a common focal plane,
separated by 106\fdg5, the basic angle. The spacecraft rotates around an
axis perpendicular to the viewing directions of the telescopes, with one
revolution every six hours.  An area of the sky is therefore first seen by the
preceding field of view, and 106.5~min later by the following field
of view.

\begin{figure}
	\begin{center}
	\includegraphics[width=\columnwidth]{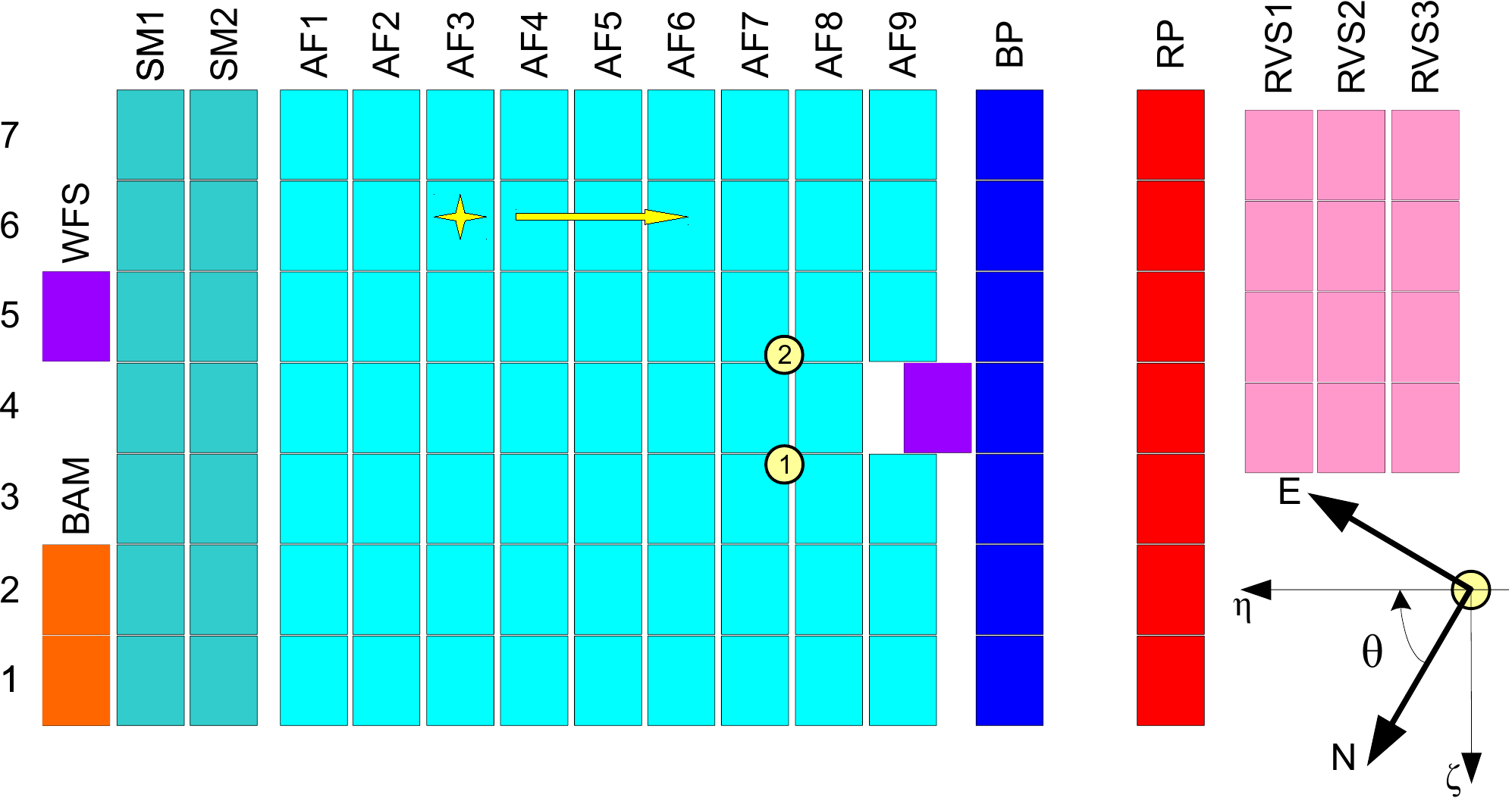}
	\caption{\label{fig:fpa} Gaia focal plane with its 106 CCDs arranged in 
	seven rows. Because of the rotation, stellar images slowly drift along the focal 
	plane from left to right. All rows contain sky mapper (SM) CCDs for detecting 
	incoming objects, eight or nine CCDs for astrometric measurements (AF1,...,AF9), 
	and CCDs for blue (BP) and red (RP) spectrophotometry. 
	In addition, some rows also include CCDs for the basic-angle monitor (BAM), for 
	wavefront sensors (WFS), and for the radial-velocity spectrometer (RVS). 
        The orientation of field angles $(\eta, \zeta)$ is also shown, and
        their origin for each of the two telescopes (yellow circles~1 and~2 for
the preceding and following fields of view, respectively). The angle
        $\theta$ exemplifies the instantaneous orientation of the focal-plane with respect
        to the equatorial coordinate system (ICRS) on the sky for one of the telescopes at a random instant of time.   
         }
	\end{center}
\end{figure}

The layout of the focal plane is shown in Fig.~\ref{fig:fpa}. The more than one
hundred  charge coupled devices (CCDs) are organised in seven rows, each with 
its own, autonomous
control unit. Owing to the rotation of the spacecraft, the stellar images will
drift over the focal plane in the along scan (AL) direction, cf.\
Fig.~\ref{fig:fpa}, in about 1.5 minutes, adjusted to coincide with the clock
rate of the CCD readout. The CCDs thus work with time-delayed integration
(TDI), and the time for shifting one line of pixels\footnote{In Gaia jargon, a
row of CCD pixels is called a line in order to avoid confusion with the
seven rows of CCDs.}, a little less than 1\,ms, is referred to as a TDI1
period.

The first two vertical strips of CCDs, denoted SM1 and SM2, have the role of sky mappers.
They have baffles that ensure they only see one telescope each, and are read in
full imaging mode. A quick onboard image analysis then produces a list of
detected, point-like objects for observation in the following CCDs.  These CCDs
see both fields of view, and only small windows around the predicted
positions of each object are actually read out.
To reduce the noise, the windows are
binned so that only one sample is obtained for each TDI1 for each window, 
rendering the windows a one-dimensional string of
samples. 
The brightest objects ($G<13$\ mag) are exempt from this binning and
therefore have two-dimensional windows. The number of
simultaneous observations has a limit, so in very dense areas the faintest
detections will not lead to an observation in every scan.

In the CCDs of the astrometric field, named AF1,..,AF9 in Fig.\ref{fig:fpa},
the fundamental observational quantities are the observation time for the
crossing of the stellar image over an imaginary line in each CCD, and the image
flux expressed in e$^-$/s.

In order to avoid or at least reduce saturation of the images, CCD features
known as gates are available at a set of CCD lines. When a gate is
activated, the charge reaching that point is blocked from progressing, so the
integration is essentially reset, and the integration time is reduced
correspondingly. This facility is used for stars brighter than about 12~mag.
An active gate affects all windows in the CCD crossing that line during the
brief moment of activation, and some windows therefore end up with multiple
gates and will be discarded if severely affected. This will remain so also
for future data releases.

After the astrometric field, the images cross the blue and red photometers
(BP and RP),
where prisms give low resolution spectra. Colours derived from these spectra
are used in the daily monitoring of the instrument and will eventually be
used when choosing the optimal point spread function (PSF) for the analysis of the astrometric 
observations, cf.\ Sect.~\ref{sec:lsfpsf}. 
Finally, the images reach the radial-velocity spectrometer (RVS), where we
get high-resolution spectra for the brighter stars. 

The spin axis is precessing, and the images therefore not only move along the
CCD columns in the AL direction, but they also have a small component
in the across-scan (AC) direction. This motion can reach 4.5~pixels for the
transit of a full CCD and therefore produces a significant AC smearing.
As most samples are binned in the AC direction, the net effect on the 
observations is small.

As mentioned, only small windows are acquired around each observed source.  In
the astrometric field they measure 12~pixels (2\farcs 1) AC, and 12--18~pixels
(0\farcs 7--1\farcs 1) AL.  As the twelve pixels in each line are binned during
readout, it is unavoidable that conflicts arise between windows for components
of double stars or in dense areas of the sky. The adopted solution is that the
brighter source gets a normal window, whereas the fainter one gets a sometimes
heavily severed (truncated) window. These latter windows have not been
processed for Gaia-DR1, which therefore generally does not contain close binaries.

It is essential for the astrometry that the angle between the two telescopes
remains very stable at a timescale of a few spacecraft revolutions, i.e.\
about a day.  Gaia is therefore equipped with a special device, the basic-angle
monitor (BAM), as further described in Sect.~\ref{sec:bam}.

\section{Daily and cyclic processing}

The Gaia science data are processed many times. First in a daily pipeline,
as detailed below, and later in several iterative large scale processes.

\subsection{Overview of the daily pipeline\label{sec:daily}}

It was realised from the start that, in a complex mission like Gaia, not only
the spacecraft but also the payload must be monitored closely on a daily
basis in order to catch any minor or major issue at an early stage.

The science telemetry is therefore treated as soon as it reaches the European
Space Astronomy Centre (ESAC) near Madrid, which houses the central hub of the
Gaia processing centres \citep{2007ASPC..376...99O}. The first process is the 
`mission operations centre interface task' (MIT), which reconstructs the telemetry
stream, identifies the various data packets, and stores them in a database.

In the daily pipeline, MIT is followed by the `initial data treatment' (IDT).  This
initial process includes reconstructing all details for each window, like its
location, shape, and gating, and the calculation of image parameters
(observation time and flux), and preliminary colours. The detailed calibration
of the image parameters is, however, not part of the pre-processing. Also the
spacecraft attitude is reconstructed with sufficient accuracy that the observed
sources are either identified in a source catalogue (Sect.~\ref{sec:xm:igsl})
during a cross-match or added as new ones. 

The outcomes of the IDT system are immediately subjected to the so-called First Look system,
which also runs within the daily pipeline at ESAC. It aims at an in-depth verification of the instrument health on board and of the scientific quality and correct on-ground treatment of the data. It produces a vast number of summary diagnostic quantities, as well as most of the daily calibrations described in Sect.~\ref{sec:ipd} and the second on-ground attitude determination described in Sect.~\ref{sec:oga}.

For similar reasons, an independent (partial) daily pipeline runs at the 
data processing centre of Turin
\citep{2012SPIE.8451E..0EM} for the scientific verification of some of the outputs from the main
ESAC pipeline which are particularly relevant for the astrometric error budget 
(Sects.~\ref{sec:valid:bamavu} and~\ref{sec:valid:aimavu} ).

\subsection{The cyclic processing\label{sec:cyclic}}

Gaia-DR1 is based on the image parameters from the daily pipeline.
However, in the future, as the data reduction progresses, new calibrations
(PSF, source colours, geometry of the instrument, etc.) will be determined, and
many elements of the pre-processing will have to be repeated within the
cyclic processing framework. 

The observations will then enter a large, iterative scheme, where better
calibrations lead to better image parameters, which again lead to improved
astrometric and photometric solutions, leading to better calibrations, etc.
Many processes participate in parallel in this scheme, and are executed once
over the whole data set in each data reduction cycle.

The cross-match and source list generation is also repeated in these cycles, in
order to better distinguish spurious from genuine detections, and to process
all detections in a coherent manner. Although a full cycle has not yet been
executed, it is such a `cyclic' style cross-match that forms the basis for the
present data release, as described in Sect.~\ref{sec:xm}.

The pre-processing corresponding to the cyclic Gaia treatment is carried out 
at the Barcelona data processing
centre \citep[][Appendix A]{jcPhd}, using the Barcelona Supercomputing 
Centre\footnote{See \url{https://www.bsc.es/marenostrum-support-services/}}.

%% file: attitude.tex
\section{Initial attitude determination \label{sec:oga}}

% Contact persons for this section: JP

The orientation of the Gaia instrument in the celestial reference frame, and
thereby the pointing direction of each telescope, is given by the attitude. For
a detailed discussion, see \cite{2012A&A...538A..78L}.  We need to know the
instantaneous pointing to the level of 100~milli-arcseconds (100~mas) in order
to safely identify the observed sources.  An early step in the daily pipeline
is therefore the reconstruction of the attitude, leading to the so-called 
first on-ground attitude (OGA1).  We use, as a starting point, the
onboard raw attitude, which is based on a combination of star tracker readings
and spin rate measurements from the astrometric observations. The raw attitude
has an offset of 10--20\arcsec, depending on the star tracker calibration, and
this offset can vary by a few arcsec during a revolution.  

The principle of the attitude reconstruction is quite simple. We use the
concept of field angles \citep{2012A&A...538A..78L}, i.e. the spherical
coordinates on the sky relative to a reference direction in each field of view,
as illustrated in Fig.~\ref{fig:fpa}, see also Sect.~\ref{sec:xm:coord}.  We
take observations of bright sources (8--13~mag) acquired with two-dimensional
windows and the field angles for each of their CCD transits.  These field
angles can then be compared to the ones computed from a specially prepared
Attitude Star Catalogue using the raw attitude at the observation time of each
transit. This comparison directly gives us the corrections to the attitude. The
major error contribution in this process comes from the quality of the star
catalogue, but this problem will disappear as soon as Gaia-based positions can
be introduced.  The attitude reconstruction is carried out before the final
image parameters for each CCD transit have been derived, and we therefore use
simple image centroids calculated with a Tukey bi-weight algorithm
\citep{tukey1960survey}. 

The  Attitude Star Catalogue in use was made by combining seven all-sky
catalogues and selecting entries based on magnitude, isolation, and astrometric
precision criteria:
\begin{enumerate}
\item The star is in the Gaia broad-band magnitude range $ 7.0 < G < 13.4$;
\item The star is isolated, e.g. it has no neighbour within 40\arcsec\ and 2 magnitudes;
\item The star is not in the Washington Double Star or Tycho Double Star
catalogues;
\item The star has positional astrometric precision better than 0\farcs 3.
\end{enumerate}
The catalogue has 8\,173\,331 entries with estimates of the positions at epoch
2000, proper motions and magnitudes (Gaia $G$, Gaia \grvs, red $R_F$ \& blue
$B_J$).  It is publicly available from the Initial Gaia Source List
\citep{2014AA...570A..87S} web-site\footnote{\url{http://igsl.oato.inaf.it/}}. 
The positions are mostly from UCAC4 \citep{2013AJ....145...44Z}.

\begin{figure}
  \begin{center}
    \includegraphics[width=\columnwidth]{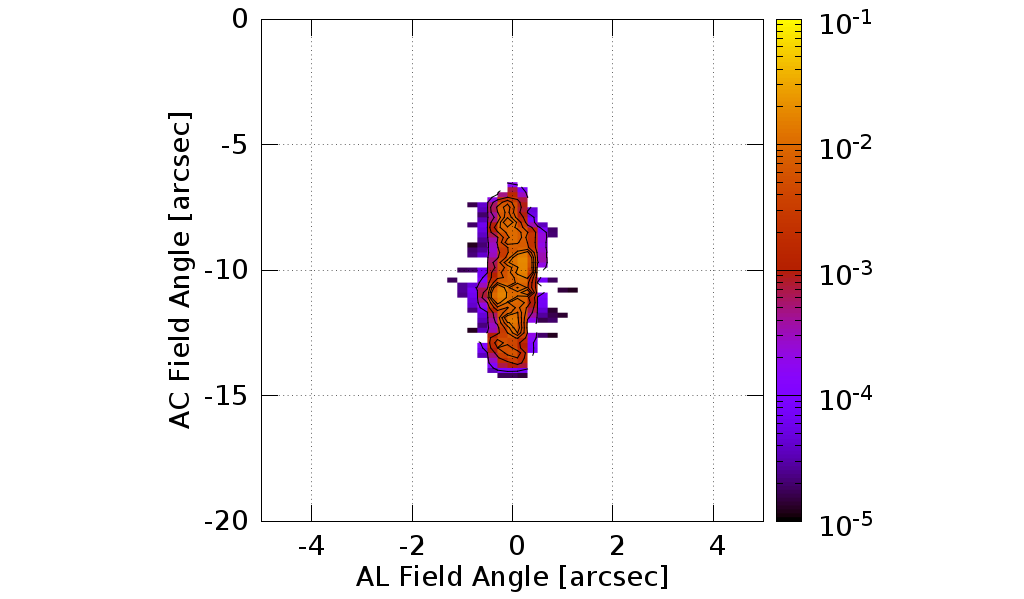}
    \caption{\label{fig:oga1xm}
      Two-dimensional histogram with the OGA1 cross-match results for one
      field of view and about one day of mission, showing the angular distance
      between the transits selected for attitude correction and their catalogue
      references in the AL, and AC directions.
      The offset and variations are largely due to small instabilities of
      the star tracker and therefore completely harmless.
    }
  \end{center}
\end{figure}

In principle the star tracker calibration may change, and we therefore carry
out the identification of the bright observations in two steps.  We first look
for any catalogue source within a very generous 30\arcsec\ of each detection,
using the initial attitude, and express the positional deviation in its AL
and AC components for each field of view. For time frames of about
half an hour, we then take the median deviations, but only among detections
with few possible identifications. In a second step, we reduce the match radius
to a still generous 5\arcsec, but centred on the median deviation. Excluding
ambiguous cases, we end up with a set of reliable matches.
Misidentifications may of course occur, and it is therefore important to have a
large number of observations, so that a few outliers do not disturb the final
solution.  Figure~\ref{fig:oga1xm} illustrates the final positional deviations
for runs over a whole day of data.

Field angles determined from the individual measurements and those from the
catalogue stars are provided to an extended Kalman filter 
\citep{LL:DMP-001}, doing first a forward and then a backward run on these
time-sorted inputs in order to minimise possible spikes on the edges.  The
result is a sequence of attitude quaternions\footnote{ See Appendix~A in
\cite{2012A&A...538A..78L} for a discussion of quaternions.}, one per valid
CCD transit (typically about 10 to 20 per second), giving the refined attitude
at each observation time. 

By comparing this refined attitude with the initial (raw) attitude we get the
attitude correction as determined by the Kalman filter, which can be expressed
in the form of differential AL and AC field angles.  A smooth cubic spline
is determined for these, acting as a reference correction, which allows running
a final consolidation step of OGA1.  This is done by detecting spikes,
that is, quaternions that diverge too much from the reference correction (more
than 0\farcs 5 AL or 1\farcs 5 AC).  Such spikes are replaced
by the reference correction, which leads to a more robust attitude
reconstruction. The smooth rotation of the spacecraft may suffer disturbances
from internal micro-clanks, but these are very small, or from external
micro-meteoroid hits, but if these reach arcsecond levels, the observations
are interrupted and the attitude reconstruction discontinued.

\begin{figure}
  \begin{center}
    \includegraphics[width=\columnwidth]{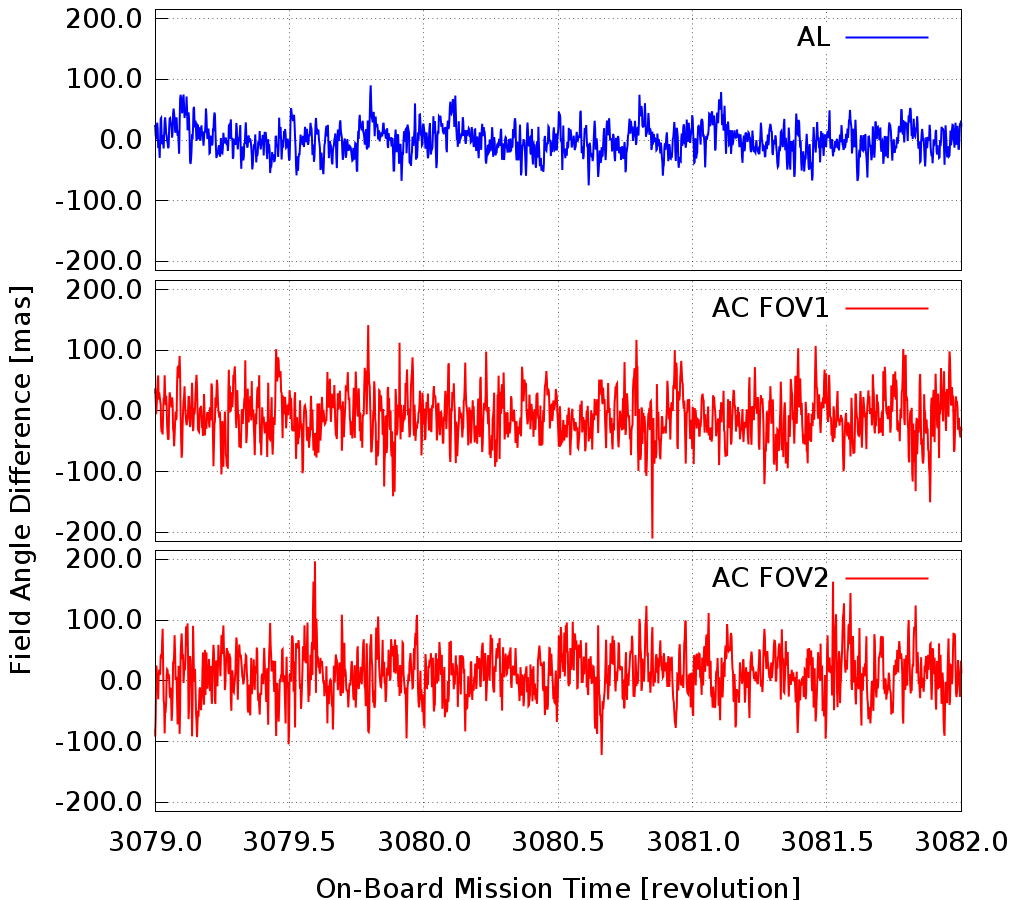}
    \caption{\label{fig:oga1vsOga2}
      Comparison of the first and second on-ground attitudes for the $\eta$, 
      $\zeta_1$ and $\zeta_2$ field angles
      (AL and AC) for 18~hours of mission data, where the indices~1 
      and~2 refer to the two Gaia telescopes (i.e.~fields of view).
    }
  \end{center}
\end{figure}

Later in the daily pipeline, the one-day astrometric solution \citep{LL:SJ-009}
produces a refined attitude determination, the second on-ground attitude
(OGA2), reaching sub-milliarcsecond precision.  Figure~\ref{fig:oga1vsOga2}
shows a comparison of the first and second on-ground attitudes, showing that
OGA1 is accurate to about 0\farcs 1. 

% ----------------------------------------------------------------------------

%% file: ipd_intro.tex
\section{Image parameter determination\label{sec:ipd}}

A main goal of the pre-processing is the determination of image parameters
for each of the several observation windows of each transit. 
We will in this section
outline the principal steps in getting from the raw spacecraft telemetry to the basic astrometric and photometric quantities 
(observation times and fluxes) for the AF and SM windows.

First of all, the window samples and the relevant circumstances like the
shape and position on the CCD of each window, the gating, any charge injections
within or preceding the window, etc., must be extracted from the various
elements of the telemetry stream. This is a complex and delicate process,
but in principle it only needs to be done once. As it is completely Gaia
internal, the details are beyond the scope of the present paper.  

The subsequent steps include determination of CCD electronic bias, background,
and a source colour, before proceeding to the image fitting itself. At this
stage of the mission, all sources are assumed to be point-like, thus close
double stars will not be processed reliably. 

\begin{figure}
  \begin{center}
    \includegraphics[width=0.95\columnwidth]{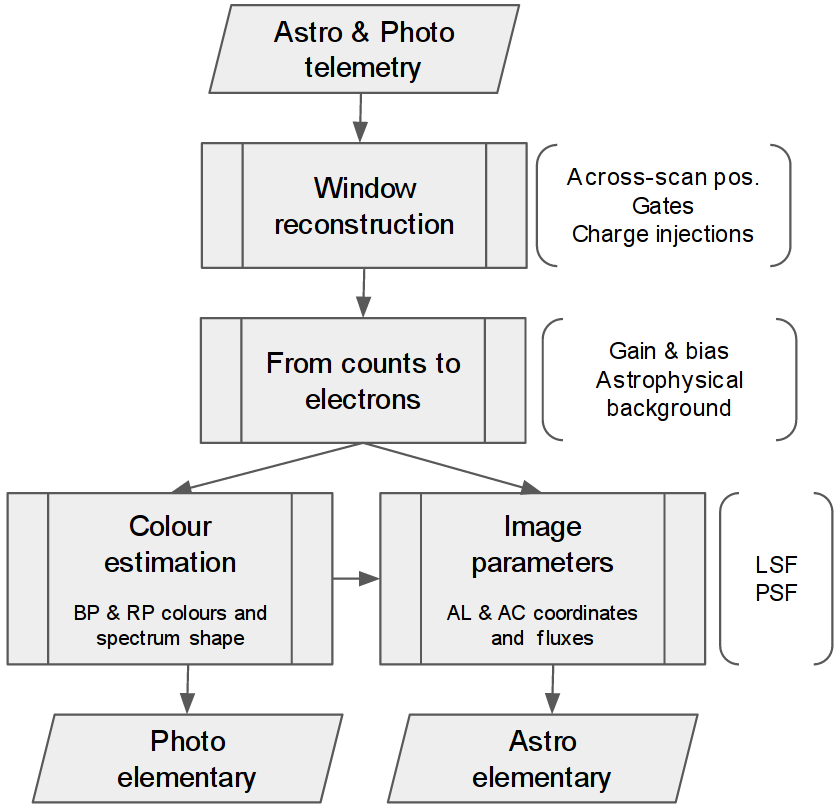}
    \caption{\label{fig:ipdFlow}
      Image parameter determination flow diagram summarising how the location within the window and the flux are obtained. The image parameters are stored as
{Astro Elementary} data and the colour parameters as {Photo Elementary}.
    }
  \end{center}
\end{figure}

Figure~\ref{fig:ipdFlow} summarises the image parameter determination process.
From each reconstructed observation, the raw CCD sample values are converted to
photo-electron counts by using the adequate gain\footnote{The gain was
calibrated before launch for each CCD module, and is typically 3.9 $\rm{e}^-/\rm{ADU}$.} after
subtracting the bias. At this stage, samples affected by CCD cosmetics,
saturation, etc.\ are masked and ignored in the following processes. We also
make sure in each window, to only use samples with the same shape and exposure
time, as truncation or gating due to a nearby source may affect some part of
the window. If the masking leaves only few samples as valid, the whole window
is discarded.

From the astrometric windows, a preliminary estimation of the image parameters
is obtained, using again a Tukey bi-weight centroiding algorithm. These
preliminary location estimations are combined with the attitude in order to
estimate the source location in the BP and RP windows,
i.e.\ the position of certain reference wavelengths. This allows us to derive
the source colour, which in principle is needed for selecting the adequate PSF
(or LSF\footnote{Line spread function, the one-dimensional equivalent of a PSF,
used for windows without AC resolution.}) for the final image fitting
(Sect.~\ref{sec:ipd:mll}).

In the daily pipeline, the requirement always to be up-to-date, in order to
monitor the instrument in almost real time, sometimes led
to postponing the image parameter determination for observations fainter
than 13~mag to the cyclic processing. This could happen due to downtime or
due to a heavy load when scanning close to the Galactic plane. Therefore around
10\% of the observations for these fainter sources did not enter Gaia-DR1.

%% file: ipd_instrument.tex
\subsection{Instrument models\label{sec:ipd:models}}

We describe below the main components of the instrument model of relevance to
the image parameters. Major effects, like bias and background, are evidently
taken into account, while some more exotic effects are briefly discussed, but
not implemented in the pipeline yet.

\input{electronic_bias}
\input{ccd_health}

\input{background}

\input{lsfpsf}
\input{cti}

%% file: electronic_bias.tex
\subsubsection{Electronic bias}
\label{sec:electronic_bias}

\begin{figure}
\centerline{\includegraphics[scale=0.5]{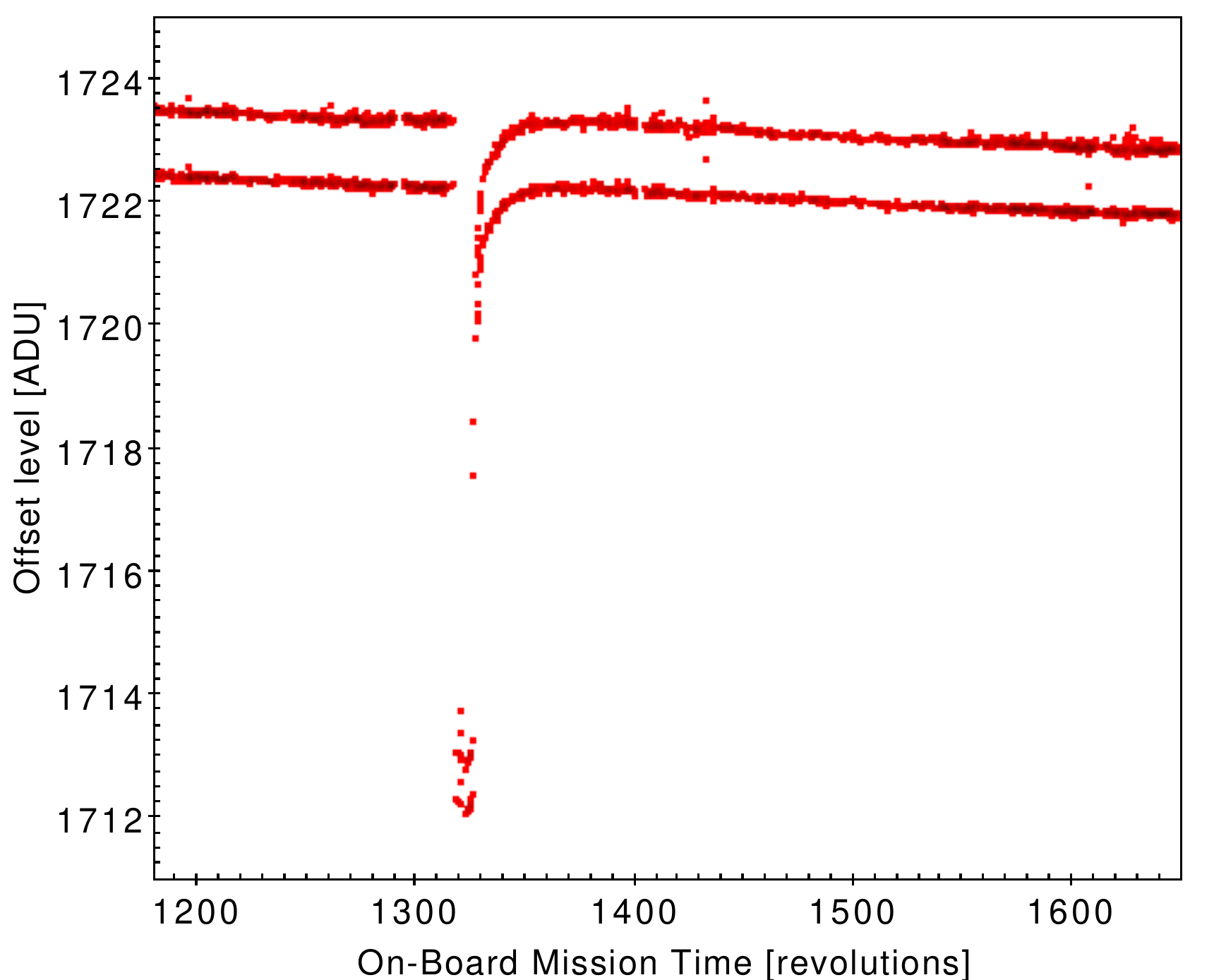}}
\caption[]{Electronic offset level in AF2 on row 4 of the Gaia focal plane.
Mean values of the approximately hourly bursts of pre\-scan data are
plotted in red.
The upper locus is for hardware-binned CCD samples containing 12~pixels AC, while
the lower locus is for unbinned data. The dip in offset level near~1320
revolutions resulted from an onboard electronic disturbance caused by
activation of payload module heaters during a decontamination period. 
One revolution of Gaia takes
6~hours and so the $x$-axis covers roughly 118~days from 19th August~2014 to
15th December~2014. One ADU corresponds to 3.9 $\rm{e}^-$.
\label{fig:electronic-offset-stability}}
\end{figure}

In common with all imaging systems that employ CCDs
and analogue-to-digital converters (ADCs) the input to the initial
amplification stage of the latter is offset by a small constant voltage to
prevent thermal noise at low signal levels from causing wrap-around across zero
digitised units (analogue-to-digital units (ADUs)).
The Gaia CCDs and associated electronic controllers and amplifiers are
described in detail in \cite{2012SPIE.8442E..1PK}. The readout registers of
each Gaia CCD incorporate 14 pre\-scan pixels (i.e.~those having no corresponding
columns of pixels in the main light-sensitive array). These enable monitoring
of the electronic bias levels at a configurable frequency and for configurable
AC hardware sampling. In practice, the acquisition of pre\-scan data is limited
to the standard unbinned (1~pixel AC) and fully binned (12~pixel AC) 
and to a burst of~1024 samples each once every 70~minutes in order that the
volume of pre\-scan data handled on board and telemetred to the ground does not
impact significantly on the science data telemetry budget. This level of
monitoring is suitable for characterising any longer timescale drifts in the
electronic offsets. For example, in Fig.~\ref{fig:electronic-offset-stability}
we show the long-time stability of one device in the Gaia focal plane. In this
case (device AF2 on row~4) the long term drift over more than
100~days is~$\approx1$~ADU apart from the electronic disturbance near
OBMT\footnote{On-board mission timeline. For convenience, the OBMT is often
expressed in units of six hours, called {revolutions} because one Gaia
spacecraft revolution has a duration close to six hours. Procedures exist to
compare OBMT to UTC to allow a transformation to TCB, which is needed for
consulting ephemerides \citep{2015jsrs.conf...55K}.} revolution~1320 
(this was caused by payload module
heaters being activated during a decontamination period in September~2014).
The approximately hourly monitoring of the offsets via the pre\-scan data
allows the calibration of the additive signal bias early in the near real-time processing
chain including the effects of long-time drift and any electronic disturbances
of the kind illustrated in Fig.~\ref{fig:electronic-offset-stability}. The
ground segment receives the bursts of pre\-scan data for all devices and distils
one or more bursts per device into mean levels along with dispersion statistics
for noise performance monitoring. Spline interpolation amongst these values is
used to provide an offset model at arbitrary times within a processing period.

Figure~\ref{fig:electronic-offset-stability} illustrates the small offset
difference between the unbinned and fully binned sample modes for the device in
question. In fact there are various subtle features in the behaviour of the
offsets for each Gaia CCD associated with the operational mode and electronic
environment. These manifest themselves as small (typically a few ADU for
non-RVS video chains, but up to $\approx$100~ADU in the worst case for RVS
devices), very short timescale ($\approx$10~$\mu$s) perturbations to the
otherwise highly stable offsets. The features are known collectively as `offset
non-uniformities' and require a separate calibration process and a correction
procedure that involves the on-ground reconstruction of the readout timing of
every sample read by the CCDs. This procedure is beyond the time-limited
resources of the near real-time daily processing chain and is left to the
cyclic data reductions at the data processing centres associated with each of
the three main Gaia instruments. However the time-independent constant offset
component, resulting from the pre\-scan samples themselves being affected and
yielding a baseline shift between the pre\-scan and image-section offset levels,
is corrected. This is achieved by measuring the effect during special calibration
periods that employ permanent activation of the gates (Sect.~\ref{sec:inst}) to
hold back photoelectrons and hence enable separation of the bias
signal from the background. The offset level difference between the pre-scan
and image-section samples is a single time-independent scalar value for each device. 
We note that while there is a post-scan pixel after the image section in the Gaia
CCD serial registers, post-scan measurements are not routinely acquired on board
nor transmitted in the telemetry. This baseline offset correction to the gross electronic bias
level of~1400 to~2600~ADU varies in size from~$-4$~ADU to~$+9$~ADU amongst the
SM, AF, BP and~RP devices. Otherwise, readout timing-dependent offset
non-uniformities are not corrected in this first Gaia data release, but
they will be corrected in cyclic reprocessing for subsequent data releases.

%% file: ccd_health.tex
\subsubsection{CCD health}
\label{sec:ccd_health}

The focal plane of Gaia contains 106 CCDs each with 4494 lines and 1966 light-sensitive 
columns, leading to it being called the `billion pixel camera'. The pre-processing requires 
calibrations for the majority of these CCDs, including SM, AF, and BP/RP, in order to 
model each window during image parameter determination. Where effects cannot be 
adequately modelled, the affected CCD samples can be masked and the observations 
flagged accordingly. The CCDs are affected by the kind of issues familiar from 
other instruments such as dark current, pixel non-uniformity, non-linearity, and 
saturation \citep[see][]{2001sccd.book.....J}. However, due to the operating principles used by 
Gaia such as TDI, gating, and source windowing, the standard 
calibration techniques need sometimes to be adapted. The use of gating generally demands multiple 
calibrations of an effect for each CCD. In essence each of the gate configurations 
must be calibrated as a separate instrument. 

An extensive characterisation of the CCDs was performed on ground, and these 
calibrations have been used in the initial processing. The effects must 
be monitored and the calibrations redetermined on an on-going basis to identify changes, 
for instance the appearance of new defects such as hot columns. To minimise disruption 
of normal spacecraft operations, most of the calibrations must be determined from 
routine science observations. Only a few calibrations demand a special mode of operation, such 
as offset non-uniformities and serial charge transfer inefficiency (CTI) measurement (see Sects.~\ref{sec:electronic_bias} and \ref{sec:cti}). 
There are two main data streams used in this calibration: two-dimensional science windows and `virtual objects'. 
The two-dimensional science windows typically contain bright stars, although a small fraction of faint stars which would otherwise be 
assigned a one-dimensional window are acquired as two-dimensional (known as `calibration faint stars'). 
Virtual objects are `empty' windows which are interleaved with the detected objects, when 
onboard resources permit. By design the virtual objects are placed according to a fixed 
repeating pattern which covers all light-sensitive columns every two hours, ensuring 
a steady stream of information on the CCD health. The virtual objects allow monitoring of the 
faint end of the CCD response while the two-dimensional science windows allow us to probe the 
bright end.

The dark signal (or dark current) is the charge produced by each column of a CCD when it 
is in complete darkness. While such condition was achieved during the on-ground 
testing it is not possible to replicate in flight as there are no shutters on Gaia. 
The observed virtual objects and science windows must therefore be used to determine the dark 
signal for each gate setting, although these also contain background, source and 
contamination signal, bias non-uniformity, and CTI effects. A sliding frame of 
50 revolutions is used to select eligible input observations, for instance those not containing 
multiple gates or charge injections. The electronic bias (including non-uniformity) is 
subtracted from each window using the pre-determined calibration (see Sect.~\ref{sec:electronic_bias}), 
and a source mask is created via an N-sigma clipping of the 
debiased samples. The leading samples in the window are also masked to mitigate 
CTI effects. A least-squares method is then used to estimate a local background 
for the window (assumed to be a constant for each sample), and this in turn can be subtracted to 
provide a measure of the dark signal in each CCD column covered by the window. In this 
manner measures can be accumulated for each column over the 50 revolution interval, and 
then a median taken to provide a robust dark signal value.

In an ideal device there would be a linear response between the accumulated charge 
and the output of the ADC at all signal levels. In reality the response typically 
becomes non-linear at high input signals for a variety of reasons \citep[see][]{2001sccd.book.....J}. 
Although the linearity has been measured before launch, a calibration has not yet 
been implemented in the daily pipeline due to the uncertainty in determination of the 
input signals, which require detailed knowledge of a range of coupled CCD effects. 
In the meantime a conservative linearity threshold has been used to allow masking 
of samples which may be within the non-linear regime. A related topic is the pixel 
non-uniformity which represents the variation in sensitivity across a CCD. In Gaia 
we observe only the integrated sensitivity of the pixels within a particular gate so 
this is known as the `column response non-uniformity'. Similarly no in-flight 
calibration has yet been performed apart from the extreme case to identify dead columns. 
These are columns which appear to have zero sensitivity to illumination and can be 
found using bright-star windows. The accumulated samples for a dead column have 
a distribution which is consistent with the expected dark signal plus read noise. The 
CCDs used on Gaia have been selected for their excellent %health. 
cosmetic quality. Details on the number and strength of the CCD defects, and their 
evolution over the course of the mission so far, are presented in Crowley 
et al.\ 2016 (this volume).

At the highest signal levels various saturation effects occur on the device and within the ADC. There can be very large 
differences in the effective saturation level across a single device, or even between 
neighbouring columns, for example due to variation in the full well capacity. For 
reasons beyond the scope of this paper the saturation level can oscillate or 
jump depending on the read-out sequencing. An algorithm has been developed to 
measure the lowest observed saturation level for each gate and column to allow 
conservative masking of samples. A Mexican hat filter is applied to the accumulation 
of samples from bright-star windows to identify over-densities of data at particular 
signal levels, using analytical significance thresholds. The lowest significant 
peak is then taken as the saturation level. If no peak is found then the maximum 
observed sample for that column is used.

The calibrations discussed above are computed daily in the framework of the First-Look system (see Sect.~\ref{sec:valid:fl}) and, if they are judged to be satisfactory, 
the corresponding software libraries are subsequently used in the pipeline.

%% file: background.tex
\subsubsection{Large-scale background}
\label{sec:background}

\begin{figure}
% examples from IDT run 370, OBMT rev 1750, 9th Jan 2015 
% trim = from left, from bottom, from right, from top in `big points' unless specified otherwise (e.g. cm).
\centerline{\includegraphics[scale=0.3,trim=0 100 5 27,clip]{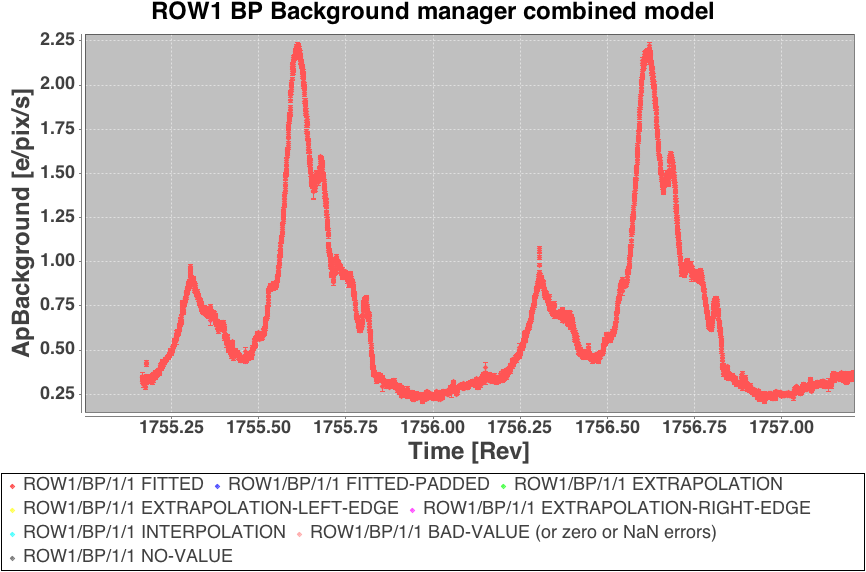}}
\centerline{\includegraphics[scale=0.3,trim=0 100 5 25,clip]{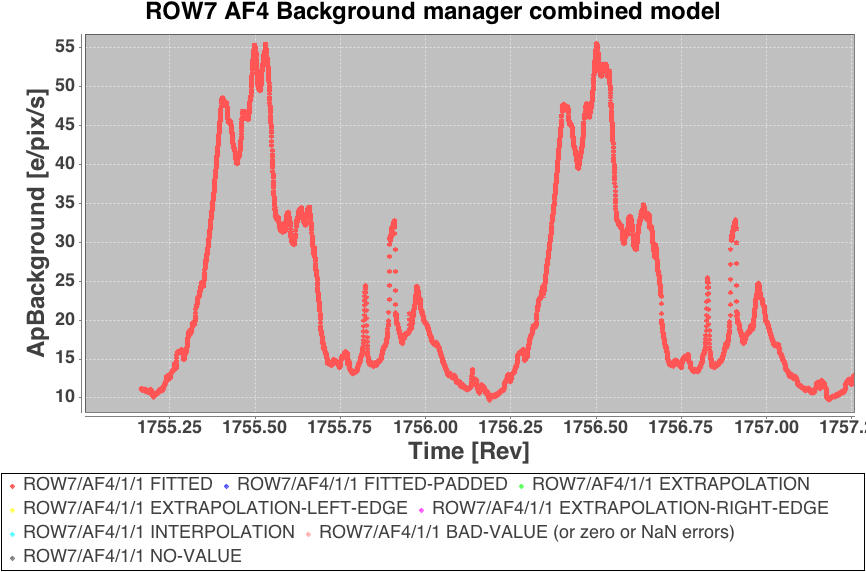}}
\caption[]{(top) Example large-scale background model in the centre of BP device on row 1 over a
period of two spacecraft revolutions in early January 2015; (bottom) the same for AF4 on row 7. The former
exhibits background levels and variations amongst the lowest over the astro/photo focal plane, while
the latter exhibits the largest.\label{fig:largescalebackgrounds}}
\end{figure}

The large-scale background signal upon which all source observations sit has several components:
i)~photoelectric background caused by 
incident photons from the diffuse astrophysical background and scattered light originating 
from the instrument itself; ii)~in the astrometric and photometric instruments, a
charge release signal following the charge injections which
are used for onboard radiation damage mitigation~\citep{2011MNRAS.414.2215P}; and
iii)~dark signal from thermal electronic charge generation within the CCD pixels (see 
Sect.~\ref{sec:ccd_health}).
As it turns out, the first of these totally dominates the others, and within the photoelectric
component it is scattered light that dominates the diffuse astrophysical background. 
Hence the dominant part of the background consists of a high-amplitude,
rapidly changing component that repeats on the satellite spin period. Furthermore, this component 
evolves slowly in both amplitude with the L2 elliptical orbital solar distance and in phase as 
the scanning attitude changes with respect to the Ecliptic and Galactic planes. Superposed on this
are transient spikes in background due to very bright stars and bright solar system objects transiting
across or near the focal plane. The approach to the daily modelling of this large-scale background, 
is to use bright-star observations to measure a two-dimensional background
surface independently for each device so that model values can be provided at arbitrary 
AL time and AC position
during downstream processing (e.g.~when deriving astrometric and photometric measurements from all
science windows). The photoelectric background signal permanently varies over several orders of magnitude
depending on instrument and spin phase with values as low as~$\approx0.1$ and up
to~$\approx50$ electrons per pixel per second; some examples are shown in
Fig.~\ref{fig:largescalebackgrounds}.

The charge release component of the large scale background appears as a relatively small periodic
modulation in those devices where charge injections are enabled for radiation damage mitigation.
An extensive discussion of the cause and effects of radiation damage in the Gaia CCDs is given in 
\cite{2011MNRAS.414.2215P}. Briefly, incident particle radiation (primarily solar protons) induce lattice
defects in the semiconductor pixels of the CCDs which manifest themselves as charge traps when
charge packets are moved through and between the pixels. These traps hold and subsequently release a certain
proportion of the passing charge cloud with characteristic timescales depending on their
physical and electro-chemical properties. The consequent CTI in
shuffling signals across the CCD pixels results in a distortion in image shape in the direction
parallel to the charge transfer. Hence we expect to see the effects of CTI in both the AL
direction (parallel to the main image section charge transfer in TDI mode) and also in the AC
direction as a result of CTI in the CCD serial 
registers. CTI effects can be mitigated in part if traps are filled by artificially
introduced charge. Hence in the main astrometric and photometric devices~(AF, BP and~RP), charge injections
are employed with duration of 4~TDI1 and repeat period of 2~s~(AF) and 5~s~(BP and~RP). The
demanded injection value is chosen so as to be large enough to introduce a useful level of charge in
all columns while at the same time avoiding saturation of fully-binned on-chip hardware samples.
In addition to the dead time introduced by injections (4 in every 2000 TDI1 in AF and in every 5000 TDI1
in BP/RP) the trade-off with the use of charge injection is an artificial inflation of the background, especially
in the lines immediately following the injection. For example, with the current level
of CTI the charge release signal in the first line after charge injection is typically between~1 and 10~electrons
per pixel per second. The signal rapidly falls to zero, however, such that by the 10th TDI line
after injection the signal is typically 1\% of this first level. Subsequently the photoelectric background signal
totally dominates for the vast majority of TDI lines between each injection event.

The daily calibration of the large-scale background is followed by a determination
of the charge release background. Sample residuals of the large-scale background, folded by
distance from last charge injection are input into a one-day calibration and a library of
charge release calibrations created. A separate calibration library for charge injection monitoring
is also produced daily. This also allows the charge release calibration to be made as a 
function of charge injection level, which is important as the AC injection profile
shows large variations from column to column. The charge injection and charge release calibrations
are made each day in order to follow the expected slow evolution of CTI. We note that the simple
decomposition of large-scale background and periodic charge release signature cannot distinguish
long-timescale charge release from the photoelectric background signal.
However, at least for the purpose of the daily pipeline,
all that is required is an empirical model with which to correct source samples, and the
sum of the background components derived above is an accurate model for this.

%% file: lsfpsf.tex
\subsubsection{Point Spread Functions and Line Spread Functions}
\label{sec:lsfpsf}

Two of the key Gaia calibrations are the LSF and PSF.
These are the profiles used to determine the image parameters for each window
in a maximum-likelihood estimation (Sect.~\ref{sec:ipd:mll}, specifically the
AL image location -- i.e.\ observation time -- source flux, plus AC location in case of
two-dimensional windows). Of these the observation time is of greatest importance in the
astrometric solution, and this is reflected in the higher requirements on the
AL locations when compared to the AC locations
\citep[][Sect.~3.4]{2012A&A...538A..78L}, where they are a factor 10 more
relaxed.  
For the majority of windows, which are 
binned in the AC direction and observed as one-dimensional profiles, an LSF is more
useful than a two-dimensional PSF.  

The PSF is often understood as the
response of the optical system to a point impulse, however in practice for Gaia
it is more useful to include also effects such as the finite pixel size, TDI
smearing and charge diffusion. This leads to the concept of the
{effective} PSF as introduced in \citet{2000PASP..112.1360A}. Pixelisation
and other effects are thereby included directly within the LSF/PSF profiles.
Calibration of the LSF/PSF is among the most challenging tasks in the overall
Gaia data processing, due to the dependence on other calibrations, such as the
background and CCD health, and due to uncertainties in crucial measured inputs
like source colour. This calibration will also become more difficult as radiation
damage to the detectors increases through the mission, causing a non-linear
distortion. Discussion of these CTI effects can be found in
Sect.~\ref{sec:cti}. Here we will focus on the LSF/PSF of the astrometric
instrument, which in the pre-processing step is applied linearly, allowing
a more straightforward modelling. 

The LSF/PSF varies over the relatively wide field of view of each telescope
(1\fdg 7 by 0\fdg 7) and with the spectral energy distribution of
an observed source. As previously discussed, the observation time of a source
depends on the gate used and, since the LSF/PSF profile can vary along even a
single CCD, all gate configurations must be calibrated independently. This can
be difficult for the shortest gates due to the relatively low number of
observations available. An LSF/PSF library contains a calibration for each
combination of telescope, CCD and gate. 

Several aspects must be considered when defining a model to represent the LSF/PSF. 
Firstly, the LSF profiles must be continuous in value and derivative, and they must be non-negative. 
By definition the full integral in the AL direction is 1, thus neglecting the flux lost 
above and below the binned AC window. This AC flux loss will be calibrated as part of 
the photometric system, but not for Gaia-DR1 (Carrasco et al.\ 2016, this volume). The LSF, $L$, is normalised as

\begin{equation}
	 \int_{-\infty}^{\infty} L(u - u_0)\mbox{\,d}u = 1
\end{equation}

\noindent where $u$ is the AL coordinate and $u_0$ is the LSF origin. The origin should be chosen to be
achromatic (the centroid of a symmetrical LSF is aligned with the origin but
this is not true in general), and since image locations are measured relative
to it, it should be tied to a physically well-defined celestial direction.
However, it is not possible to separate geometric calibration from chromaticity
effects within the daily pipeline; this requires the global astrometric
solution from the cyclic processing. The origin is therefore fixed as 
$u_0 = 0$ and consequently there will be a colour-dependent bias in this
internal LSF calibration. The LSF profile is used to model the expected
de-biased photo-electron flux $N_k$ of a single stellar source, including
noise, by

\begin{equation}
     \lambda_k \equiv \mbox{E}\left( N_k \right) = \beta\tau + f\tau L(k - \kappa)
     \label{eq:winModel}
\end{equation}

\noindent where $\beta$, $\tau$, $f$ and $\kappa$ are the background level, 
the exposure time, the flux of the source, 
and the AL image location. The index $k$ is the AL location of the CCD sample under consideration. The actual photo-electron counts will include a Poissonian noise component and a Gaussian readout noise component, in addition.
There are two fitted parameters: the flux, $f$, is the basic input to the photometric processing chain, while
the AL location, $\kappa$, gives the transit time for the astrometric processing. The background 
level $\beta$ is not fitted here but is taken from the calibration described in Sect.~\ref{sec:background}.

For the practical application, the LSF can be modelled as a linear combination of basis components 

\begin{equation}
	L(u) = \sum_{m=0}^{N-1} w_m B_m(u)
\end{equation}

\noindent where $N$ basis functions are used. The value $B_m$ of each basis function $m$ at coordinate $u$
is scaled by a weight $w_m$ appropriate for the given observation. A set of basis functions 
can be derived through principal component analysis (PCA) of a collection of LSF profiles 
chosen to represent the actual spread of observations, i.e. covering all devices 
and a wide range of source colours and smearing rates. An advantage of PCA is that 
the basis functions are ranked by significance, allowing selection of the minimum 
number of components required to reach a particular level of residuals. These basis 
functions can in turn be chosen in a variety of ways; we have used a bi-quartic spline\footnote{These splines are defined in \citet[][Appendix B]{LL:GAIA-LL-046}.} 
model with a smooth transition via a fourth order polynomial to the $u^{-2}$ diffraction profile expected at the LSF wings \citep{LL:LL-084}. Further 
optimisation can be achieved to assure the correct normalisation by transforming 
these bases, although this is beyond the scope of this paper. A set of 51~basis 
functions were determined from pre-flight simulation data, each represented using 75 coefficients
\citep{LL:LL-084}. The 20 most significant functions 
have been found to adequately represent real LSFs, although further improvements 
are possible. 

With a given set of basis functions $B_m$ the task of LSF calibration becomes 
the determination of the basis weights $w_m$. These weights depend on the observation 
parameters including AC position within the CCD, effective wavenumber of the source, AC smearing and others. 
In general the observation parameters can be written as a vector $\vec{p}$ and the weights thus as $w_m(\vec{p})$. 
To allow smooth interpolation, each basis weight can be represented as a spline surface where each dimension 
corresponds to an observation parameter. In the implemented calibration system, each dimension can be configured separately with 
sufficient flexibility to accommodate the actual structure in the weight surface, i.e. 
via choice of the spline order and knots. In practice, the number of observation parameters 
has been restricted to two: AC position and effective wavenumber (i.e.~source colour) for AL LSF, and AC smearing 
and effective wavenumber for the AC LSF. The coefficients of the weight surface are formed from the outer product of two splines with $k$ and $l$ coefficients 
respectively. There are therefore $k \times l$ weight parameters per basis function which must be fitted. 

\begin{figure}
	\begin{center}
	\includegraphics[width=\columnwidth]{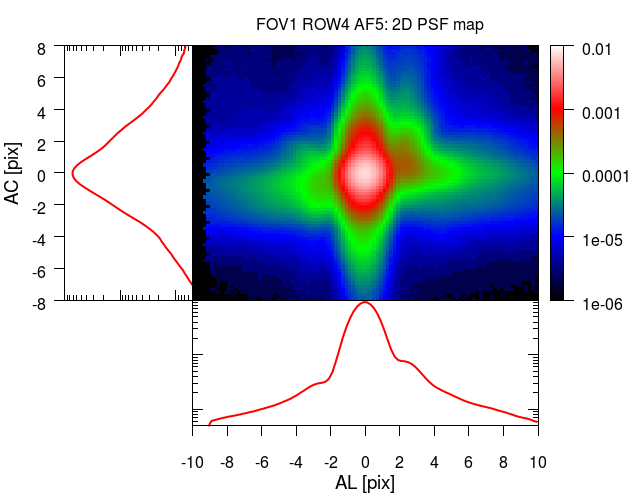}
	\caption{\label{fig:full2DPSF} Typical PSF for 
	a device in the centre of the field of view. This two-dimensional map has then been marginalised 
	in the AL and AC directions to form LSFs (left and bottom respectively). 
        The field measures 1.2\arcsec (AL) by 2.8\arcsec (AC).}
	\end{center}
\end{figure}

\begin{figure}
	\begin{center}
	\includegraphics[width=\columnwidth]{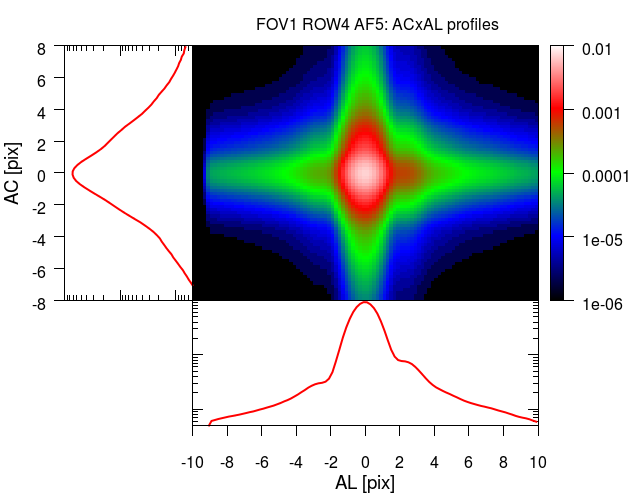}
	\caption{\label{fig:ALxACPSF} 
         Reconstruction of the PSF based on the LSFs seen in 
             Fig.~\ref{fig:full2DPSF}. 
        The PSF model used by IDT when processing two-dimensional windows 
        is the cross-product of one-dimensional AL and AC LSFs. 
        It is clear that, although the gross 
	structure is present, the asymmetry information has been lost. 
        The field measures 1.2\arcsec (AL) by 2.8\arcsec (AC).}
	\end{center}
\end{figure}

\begin{figure}
	\begin{center}
	\includegraphics[width=\columnwidth]{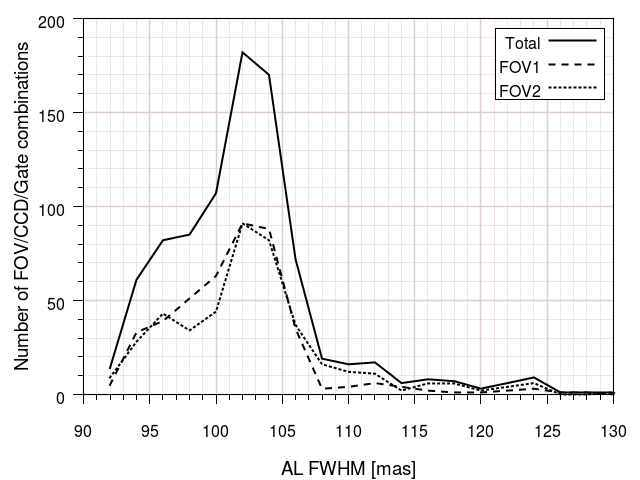}
	\caption{Histogram of the FWHM in the AL direction for all combinations of field of view, CCD and
	gate in the AF instrument at a mean colour, derived from a preliminary PSF calibration for two-dimensional windows.
	The solid black line shows the total; the dashed and dotted lines show 
        the preceding and following fields of view
%       FOV1 and FOV2 
        respectively. The median FWHM is 103 mas (1.75 pixels); in most cases the FWHM is below 108 mas,
	although there is a tail to 132 mas populated by CCDs at one corner of the
	focal plane.\label{fig:fwhms} }
	\end{center}
\end{figure}

\begin{figure}
	\centerline{\includegraphics[width=\columnwidth]{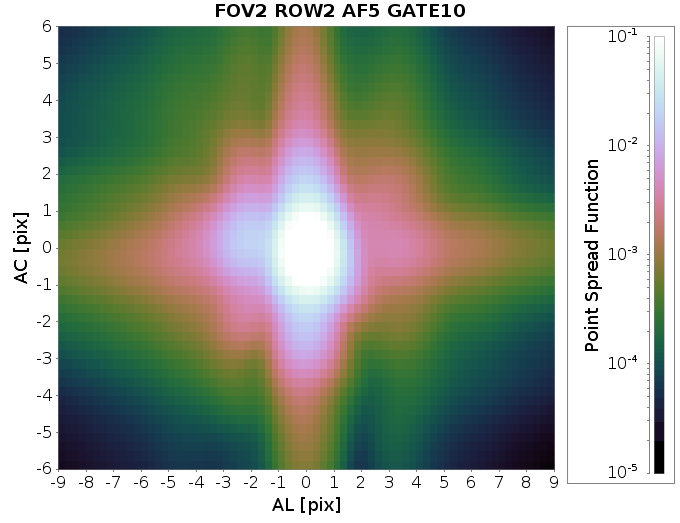}}
	\centerline{\includegraphics[width=\columnwidth]{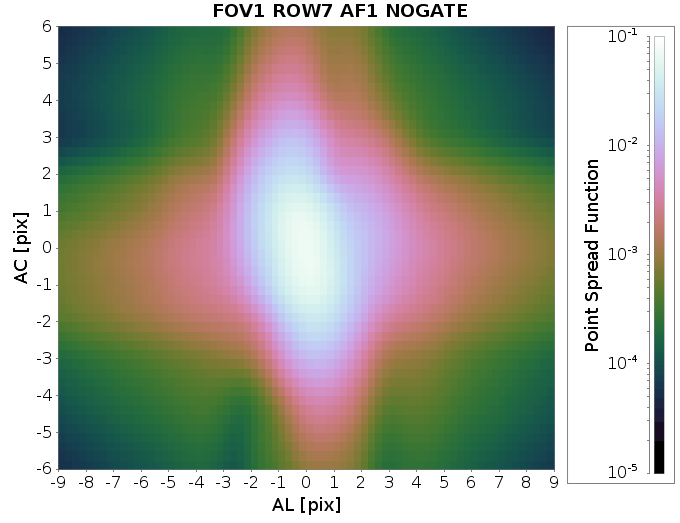}}
	\caption[]{Examples of the PSF for the devices with the smallest 
%       (FOV2 ROW2 AF5 GATE10;
	(93.1 mas) and largest 
%       (FOV1 ROW7 AF1 NOGATE; 
        (132.4 mas) AL FWHMs. The largest
	FWHMs are in devices in one corner of the focal plane (AF1/2 in rows 6 \& 7), while the smallest FWHM is
	close to the centre (AF5, row 2). The longer gates usually have larger FWHMs in the AL profile due to
	the variation in the PSF along the CCD. We also see here the effect of the greater AC smearing in the
	ungated PSF profile.\label{fig:psfExamples}}
\end{figure}

The rectangular telescope apertures in Gaia led to a simple model to approximate the PSF in the daily pipeline. The 
PSF is formed by the cross product of the AL and AC LSFs. This model has a relatively 
small number of parameters to fit at the price of being unable to represent all the structure in 
the PSF. A more sophisticated full two-dimensional model shall be available for the cyclic processing 
systems (Sect.~\ref{sec:cyclic}), where there are fewer processing constraints than in the daily pipeline. We have confirmed that the AC$\times$AL approximation does not 
introduce significant bias into the measured observation times for two-dimensional windows. 
Experiments to compare the fitted observation times using the AC$\times$AL versus a full two-dimensional PSF
indicate a systematic bias of $2.3 \times 10^{-4}$ pixels for point sources. An 
example of the PSF and its reconstruction via the AC$\times$AL model is presented in 
Figs.~\ref{fig:full2DPSF} and \ref{fig:ALxACPSF}. It is clear from these figures 
that there are asymmetric features in the PSF. The Gaia optical system uses 
three-mirror anastigmatic telescopes to minimise aberrations. However there are 
six reflectors and a degree of mirror contamination, which introduce colour-dependent PSF anisotropy 
(see also Gaia collaboration, Prusti et al.\ 2016, this volume). Hence, the PSF varies in time (with occasional refocus) 
and across the focal plane, as seen in Figs.~\ref{fig:fwhms} and \ref{fig:psfExamples}, which demonstrate the 
spread of the AL full width half maximum (FWHM) values.

LSF calibrations are obtained by selecting {calibrator} observations.
These are chosen to be healthy (e.g. nominal gate, regular window shape) and
not affected by charge injections or rapid charge release. Image parameter and
colour estimation for the observation must be successful; good bias and
background information must be available. With these data
Eq.~\ref{eq:winModel} can be used to provide an LSF measure per unmasked
sample. In the case of two-dimensional windows, the observation is binned in the AL or AC
direction as appropriate to give LSF calibrations. The quantity of data
available varies with calibration unit\footnote {A `calibration unit' in the
Gaia jargon is a given combination of window class (e.g.~one-dimensional, two-dimensional), gate, CCD,
telescope, possibly time interval (days, weeks, months), and possibly other
parameters, e.g. AC coordinate interval on a CCD.  }.  For the most common
configurations (faint ungated windows) there are many more eligible
observations than can be handled, thus a thinned-out selection of calibrators
is used. 

A least-squares method is used 
with the LSF calibrations to fit the basis weight coefficients. The Householder 
least-squares technique is very useful here as it allows calibrators to be processed 
in separate time batches and for their solutions to be merged. The merger can also be 
weighted to enable a running solution to track changes in the LSFs over time \citep[see][]{book:newhip}. 
Various automated and manual validations are performed on an updated LSF library before
approval is given for it to be used by the daily pipeline for
subsequent image parameter determination. There are checks on each solution to
ensure that the goodness-of-fit is within the expected range, that the number of
degrees-of-freedom is sufficiently positive, and that the reconstructed LSFs are
well-behaved over the necessary range of $u$. Individual solutions may be rejected and the corresponding
existing best-available solution be carried forward. In this way an operational LSF library always has a full complement of solutions for all devices and nominal configurations.

The LSF solutions are updated daily within the real-time system, although they are 
not approved for use at that frequency. Indeed, a single library generated during 
the commissioning phase has been used throughout the period covered by Gaia-DR1,
in order 
to provide stability in the system during the early mission. This library has a 
limited set of dependencies including field of view and CCD, but it does not include 
important parameters such as colour, AC smearing or AC position within a CCD. As such it is 
essentially a library of mean LSFs and AC$\times$AL PSFs. This will change for future Gaia data releases.

%% file: cti.tex
\subsubsection{Charge Transfer Inefficiency}
\label{sec:cti}
% Contact person: MDV

% Intro to Gaia CTI inc refs.
% - What is CTI --> trapping
% - What causes CTI --> mostly solar protons and ions, plus galactic and manufacturing
% - How will it affect Gaia --> systematic location and flux biases
% - Difficult to model --> TP refs etc for CDM

CTI is one of the most challenging effects to be 
calibrated in Gaia. It is caused by the presence of atomic displacements in the silicon 
lattice of the CCD, which can capture electrons while charge packets are transferred across 
each pixel \citep[see][]{2001sccd.book.....J}. Some defects are created during the manufacturing process and by % Galactic 
cosmic rays, however most are a consequence of energetic solar particles. The 
number of traps will increase gradually over the course of the mission, with some 
relatively large steps in the damage level after major solar events such as coronal mass ejections. The 
capture and subsequent release of electrons creates a distortion in the observed stellar profiles. 
Without mitigation this distortion will lead to biases in the measured image locations
and fluxes. The CTI effects are known to be difficult to model for several reasons, for instance
the non-linearity of the capture and release processes, the dependency on the previous illumination history (i.e.~the already existing 
trap occupancy), the degeneracy in the model parameters, and the continual evolution of the damage level. Much work was 
performed before launch to investigate the expected CTI response at various damage levels and possible 
mitigation strategies \citep{2012MNRAS.419.2995P,2012MNRAS.422.2786H}. The most 
promising technique is a full forward-modelling approach via a charge distortion model 
(CDM) as described in \citet{2013MNRAS.430.3078S}.   

% Current mitigation
% - Current levels of CTI
% - SBC, CIs
% - Background levels
% - Link back to CR/FPR section.

The CTI-related strategy for Gaia-DR1 has been limited 
to the onboard mitigation measures plus charge release characterisation. The accumulated 
damage is less than that predicted for this stage in the mission and the onboard 
mitigation techniques have been working well \citep{2014SPIE.9154E..06K}. These include the inclusion of a
`supplementary buried channel' \citep{2013MNRAS.430.3155S} in the CCDs to assist in the transfer of small 
charge packets, and also in the activation of charge injections. The regular injections 
fill many traps such that fewer are available to damage the science packets. A 
fortunate side-effect of the unwanted high background levels from straylight is that this 
also acts to keep traps filled. Charge captured from injections is gradually 
released to form trails as described in Sect.~\ref{sec:background}. The release 
profiles can be steep in the first few tens of TDI lines after the injection, and 
this could cause a location bias in the image parameter determination if not included in the background model.

% Future mitigation by Electronic Corrections
% - Performance limitations in IDT
% - Not yet activated due to low CTI levels
% - Updates to LSF/PSF basis weights to modify profiles as a function of CTI parameters
% - Also allows mean LSF normalisation to vary to accomodate flux loss
% - Limited to point sources
% - Interim solution while CDM is developed using mission data

Further mitigation is possible via a CDM, but certain limitations apply. The daily, near
real-time processing imposes constraints on the resources available 
for calibration and application of a CDM. In addition, the full illumination history 
of the CCDs is not available at this stage of processing, limiting the accuracy 
of the damage prediction. However, an interim model for CTI mitigation, known as the `electronic corrections',
has been developed for use in the daily chain, although it has not yet been activated due to the presently still low damage levels. This model 
attempts to capture the change in the observed source profile due to CTI by 
updating the LSF/PSF basis weights (see Sect.~\ref{sec:lsfpsf}) as a function of the parameters which most 
determine the damage level. These parameters include the time since the last charge 
injection, the source magnitude and the AC coordinate on the CCD (which strongly affects 
the amount of distortion caused by traps in the serial register). The model also 
permits variation in the mean LSF normalisation to accommodate flux loss. The 
electronic corrections should be a practical solution for mitigation of CTI for 
point sources, although a full CDM will still be required for the more complete cyclic processing.

%% file: ipd_sso.tex
\subsection{Detection of object motion\label{sec:ipd:sso}}

As mentioned in 
the introduction to this section, the preliminary centroids of the astrometric
field windows, leading to AL and AC pixel coordinates of the
detected astronomical images, are combined with the attitude to propagate the
image positions to the photometric CCDs.  A linear fit on the astrometric CCD
centroids is done for this, also requiring the geometric calibration of the
instrument.  As a by-product, the resulting fit provides an estimate of the
motion of the source on the sky, which is used as one of the indicators when
looking for solar system objects.

Because of the nature of the observations, motions can primarily be detected in
the AL direction, and motions much above 15~mas/s may be lost, partly due
to image smearing, and partly due to the objects falling outside the allocated
windows on the AF CCDs.

%% file: ipd_xp.tex
\subsection{Photometric processing of BP and RP\label{sec:ipd:xp}}

% Contact person: AB

The four Gaia instruments operate in different wavelength bands, with
SM/AF covering approximately 330--1050\,nm; BP 330--680\,nm; RP 640--1050\,nm; 
and RVS 845--872\,nm. For a detailed discussion of these bands see
\cite{2010A&A...523A..48J} or Gaia collaboration, Prusti et al.\ (2016, this volume).

As explained in the introductory part of this section, the source colours are
needed to select the appropriate PSF (or LSF) for the image parameter
determination from the AF measurements. Although the colours have not been used
in selecting the PSF/LSF for Gaia-DR1 the photometric measurements made by the
prism photometers were nevertheless processed as part of the IDT. The BP/RP
(Photo) telemetry is first turned into a preliminary BP/RP spectrum, following
the steps outlined in Fig.\ \ref{fig:ipdFlow}. That is, the mean bias and large
scale background are removed and the counts are converted to electrons using
the appropriate gain. Figure~\ref{fig:largescalebackgrounds} shows an example
of the large scale background variation encountered in one of the BP CCDs. As
for the AF window processing, samples affected by CCD cosmetics or saturation
are masked, and the truncation and gating of the samples is taken into account.
For Gaia-DR1 truncated windows were treated if the truncation affected
only the window edges and the onboard magnitude estimate for the source was
brighter than 10, otherwise the window was discarded. The measurements for
bright stars (two-dimensional windows) were converted to one-dimensional windows by summing the counts in
the AC direction, where saturated samples were accounted for by employing a
crude LSF in the AC direction and estimating the total counts by fitting this
LSF (accounting for the masked samples).

At this stage all BP/RP data are in the form of one-dimensional windows, listing the flux in 
${\rm e}^-/{\rm s}$ for
each sample. Because of the prisms in the optics path the BP/RP samples correspond to an effective wavelength.
Further processing requires the assignment of the correct wavelength to each location within the
window. This is done through estimating the position of a reference wavelength  %({\bf NOTE: for this
%DR the transit prediction for BP/RP was not active?})
% Indeed; confirmed by JP 
within the window and then applying a
dispersion curve that relates wavelength to location within the window. As a result each sample has
an assigned wavelength\footnote{
For the data entering the pre-processing for Gaia-DR1, the nominal wavelengths and dispersion curves were still used.},
allowing for the subsequent calculation of several colour parameters that
characterise the (uncalibrated) flux distribution:
\begin{itemize}
  \item The broad band fluxes for each of BP and RP are obtained by simply summing all the sample
    values in the window. The corresponding magnitudes, \gbp\ and \grp\, and the broad band colour
    \bpminrp, are calculated using a nominal magnitude zero-point.
  \item The flux in the RVS band is estimated by summing the three sample values closest to the
    centre of the RVS wavelength band (858.5~nm).
  \item The effective wavenumber, \nueff, is calculated from the sample values and the corresponding
    wavelengths according to the following formula:
    \begin{equation}
      \nueff = \frac{\sum_i^n s_i^\mathrm{BP}\lambda^{-1}_i+\sum_j^m
      s_j^\mathrm{RP}\lambda_j^{-1}}{\sum_i^n s_i^\mathrm{BP}+ \sum_j^m
      s_j^\mathrm{RP}}\,,
      \label{eq:nueff}
    \end{equation}
    where the wavelengths $\lambda_i$ and $\lambda_j$ are defined for the AL pixel coordinate
    corresponding to the middle of the samples with flux values $s_i^\mathrm{BP}$ or
    $s_j^\mathrm{RP}$. The value of \nueff\ summarises the shape of the prism spectra (i.e.\ the
    source spectral energy distribution) in one number and was shown to correlate very well with the
    chromatic shifts in the image locations induced by the PSF dependency on colour
    \citep{LL:JDB-028}. 
    As these centroid shifts vary almost linearly with the effective wavenumber,
    this quantity is preferred, over e.g.\ the wavelength, to characterise the
    spectral energy distribution. 
    The effective
    wavenumber thus forms an important input in the selection of the correct PSF/LSF for the image
    parameter determination (although this is not used for Gaia-DR1).
  \item Finally so-called spectral shape coefficients are calculated which is just the summation of
    the BP/RP fluxes over a limited set of samples, defining a pseudo wavelength band. Four
    coefficients are calculated for each of BP and RP. These provide a compact representation of
    the prism spectrum shapes, which can in the future be used in combination with \nueff\ to refine
    the selection of the PSF/LSF for the image parameter determination.
\end{itemize}

We stress that the above photometric quantities are all uncalibrated and only intended for use
within IDT. The full treatment and calibration of the Gaia photometric measurements takes place
within the dedicated photometric processing pipeline (see van~Leeuwen et al.; Carrasco et al. 2016, this volume).

It may happen that one or both of the BP and RP measurements is missing for a
given source. This can be due to issues in the onboard data collection process
or because the sample data could not be processed in IDT. In those cases a
default set of colour parameters is assigned to the source if both BP and RP
data are missing, while the colour parameters are predicted from the onboard
estimate of the $G$-band flux and the available BP/RP data if only one of BP or
RP is missing. The predictions are based on polynomial relations between the
various broad band colour combinations.  For example if RP data are missing, the
value of {\grp} is estimated from the {\bpminrp} versus $(G_\mathrm{VPU}-\gbp)$
relation -- where $G_\mathrm{VPU}$ is the onboard magnitude estimate --
while {\nueff} is estimated from the {\nueff} versus $(G_\mathrm{VPU}-\gbp)$\
relation. The
polynomial relations were derived from pre-launch simulated data and have not
been updated for Gaia-DR1.

%% file: ipd_mll.tex
\subsection{Fitting the model: image parameter determination\label{sec:ipd:mll}}

The final image parameters from the daily pipeline are determined using a
dedicated maximum-likelihood method \citep{LL:LL-078} determining the flux and
AL window coordinate, and for two-dimensional windows also the AC coordinate.  Starting
values for the iteration are obtained using the Tukey bi-weight centroiding
mentioned in the beginning of this section.  

As mentioned in Sect.~\ref{sec:lsfpsf}, only mean LSFs, established for each
CCD and field of view during the in-orbit commissioning phase, have been used
for Gaia-DR1. As a consequence, dependences on time, source colour, and image
motion AC have not been included, although they are obviously relevant. 

The formal errors of the AL coordinate, from the transit of one CCD, is around
0.06~mas for observations brighter than 12~mag, reaching 0.6~mas at 17~mag, and
3~mas at 20~mag.  However, as we have been using mean LSF/PSFs, we are still
far from utilising the full potential indicated by the signal to noise ratio,
and the goodness of fit is rather poor, and can be well above 10 for observations 
brighter than 14~mag.
The actual residuals in the astrometric solution are discussed in detail by 
Lindegren et al. (2016, this volume, especially Appendix D) and are around 
0.6~mas for a single CCD transit for the bright sources. This includes all 
unmodelled contributions from chromatic shifts, attitude disturbances, etc.

%% file: xm_intro.tex
\section{Cross-matching\label{sec:xm}}

The cross-matching (or shorter: cross-match) provides the link between the Gaia detections and the entries in the Gaia working catalogue\footnote{
This consists of the initial Gaia source list (IGSL), see Sect.~\ref{sec:xm:igsl}, plus -- in general -- the new sources created by a preceding cross-match process.}.
It consists of
a single source link for each detection, and consequently a list of linked detections for each source. When a
detection has more than one source candidate fulfilling the match criterion, in principle only one is linked,
the {principal match}, while the others are registered as {ambiguous matches}.  

To facilitate the identification of working catalogue sources with existing astronomical
catalogues, the cross-match starts from an initial source list, as explained in
Sect.~\ref{sec:xm:igsl}, but this initial catalogue is far from complete.  The
resolution of the cross-match will therefore often require the creation of new
source entries.  These new sources can be created directly from the unmatched Gaia detections. 

A first cross-match is carried out in the daily pipeline. It is mainly required for:

\begin{itemize}
\item The refinement of the attitude as explained in Sect.~\ref{sec:oga};
\item The science alerts, especially new variable sources or potentially new solar system objects \citep{2016arXiv160102827W}; 
\item The very first, global astrometric and photometric solutions prior to the first cyclic cross-match execution;
\item The daily monitoring of the instrument as explained in Sect.~\ref{sec:valid:fl}.
\end{itemize}

In the cyclic processing, the cross-match is revised using the improvements on
the working catalogue, of the calibrations, and of the censoring of spurious
detections (see below).  Additionally, a revision is needed because the daily
cross-match works on a limited data set.  Therefore, the resolution of dense sky
regions, multiple stars, high proper motion sources and other complex cases
will be deficient.  Each such revision completely replaces any previous
cross-match solution. For Gaia-DR1, a cyclic cross-match starting from the IGSL
(see below) was used.

%% file: xm_igsl.tex
\subsection{The IGSL \label{sec:xm:igsl}}

The Gaia catalogue has an astrometric accuracy level corresponding to a very
small fraction of a pixel and also a fraction of a pixel
in terms of resolution. However, the starting catalogue used for the daily
processing has been initialised with sources of a quite heterogeneous provenance,
and this provenance -- as well as its much lower angular resolution -- must be taken into account when determining the proper
source match.

\begin{figure}
\centering
\includegraphics[width=1.0\columnwidth]{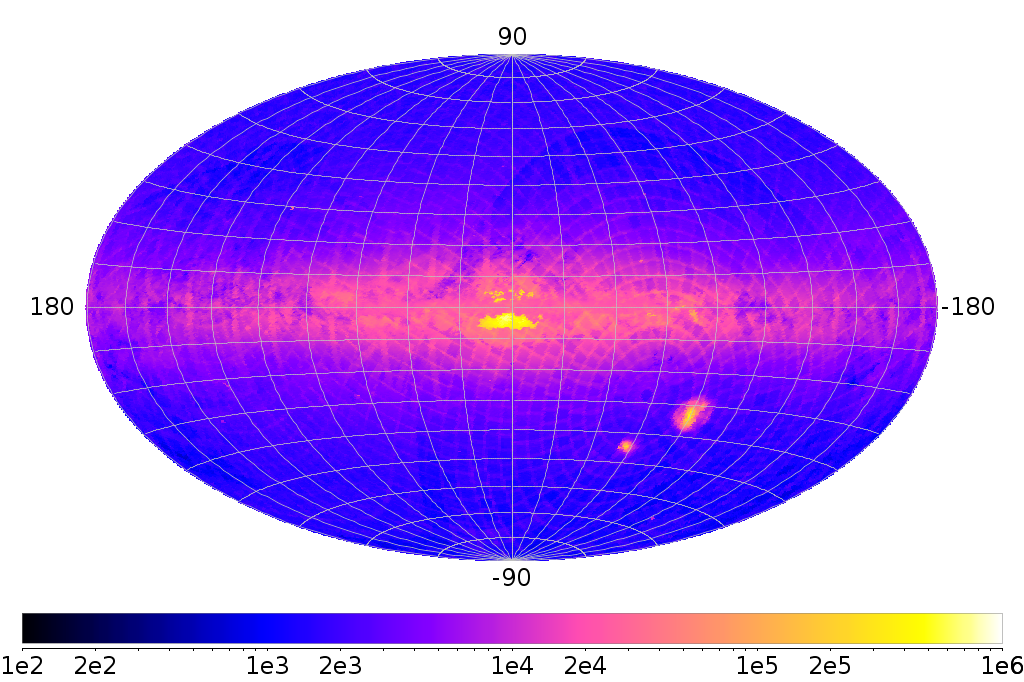}
\caption{Density of objects included in the IGSL (galactic coordinates). The
square grid structure is due to differing photometric information and
completeness in the overlap of the Schmidt plates that were used for the PPMXL
and GSC2.3 input catalogues. The bands that perpendicularly traverse the
Galactic plane are due to extra objects and photometric information from the
SDSS surveys.
\label{fig:xmIgsl}}
\end{figure}

The starting catalogue is the IGSL which was
compiled from the best optical astrometry and photometry information on
celestial objects available before the launch of Gaia: GEPC, GSC2.3, LQRF, OGLE,
PPMXL, SDSS, UCAC4, Tycho-2, Sky2000 and HIPPARCOS, as described in
\cite{2014AA...570A..87S}. This catalogue was frozen before launch and no
updating of it was foreseen for the mission. Figure~\ref{fig:xmIgsl} plots the
density of objects included in the IGSL in galactic coordinates. The IGSL has
more than 1.2 billion entries with positions, proper motions (if known) and a
blue and red magnitude, plus predicted $G$ and \grvs\ magnitudes. All IGSL
sources have been given unique source identifiers that 
incorporate a spatial HEALPix index \citep{2005ApJ...622..759G}.

%% file: xm_scene.tex
\subsection{Scene determination\label{sec:xm:scene}}

A preparatory step in the cyclic cross-match is to establish what we call the
{scene}, i.e.\ a list of known sources transiting at or within a few
degrees of the focal plane. For the cross-match, we are primarily interested in
sources that will probably not be detected directly, but still leave many
spurious detections, for example from diffraction spikes or internal reflections.
The scene is established entirely from catalogue sources and planetary
ephemerides\footnote{For Gaia-DR1 only the major planets are considered,
but later also the brighter satellites and asteroids will be included.}, and is
therefore limited by the completeness and quality of those input tables. The
scene is also useful for other cyclic processes, like the background
estimation, PSF calibration etc., but that goes beyond the scope of this paper.

%% file: xm_spurious.tex
\subsection{Spurious detections \label{sec:xm:spur}}

The Gaia onboard detection software was built to detect point-like images on the SM CCDs and to autonomously discriminate
star images from for instance cosmic rays. For this, parametrised criteria of the image shape are used, which need
to be calibrated and tuned. There is clearly a trade-off between getting a high detection probability for stars at
20~mag and keeping the detections from diffraction spikes (and other disturbances) at a minimum. A study of the detection capability, in
particular for non-saturated stars, double stars, unresolved external galaxies, and asteroids is provided by
\cite{2015AA...576A..74D}.

The main problem with spurious detections arises from the fact that they are numerous (15--20\% of all detections), and
that each of them may lead to the creation of a (spurious) new source during the cross-match. Therefore, we must classify 
detections as either genuine or spurious, and only consider the former in the cross-match.

\begin{figure}
\centering
\includegraphics[width=1.0\columnwidth,clip=]{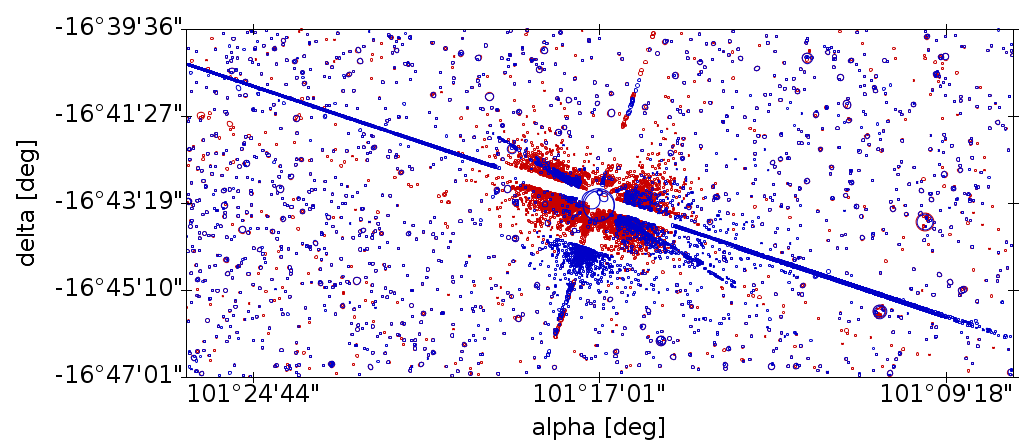}
\caption{13\,172, mostly spurious, detections from two scans of Sirius, one shown in blue and one in red. The majority of the spurious detections are fainter than 19~mag.
In the red scan Sirius fell in between two CCD rows. 
\label{fig:xmBsSpuriousExamples}}
\end{figure}

\begin{figure}
\centering
\includegraphics[width=1.0\columnwidth]{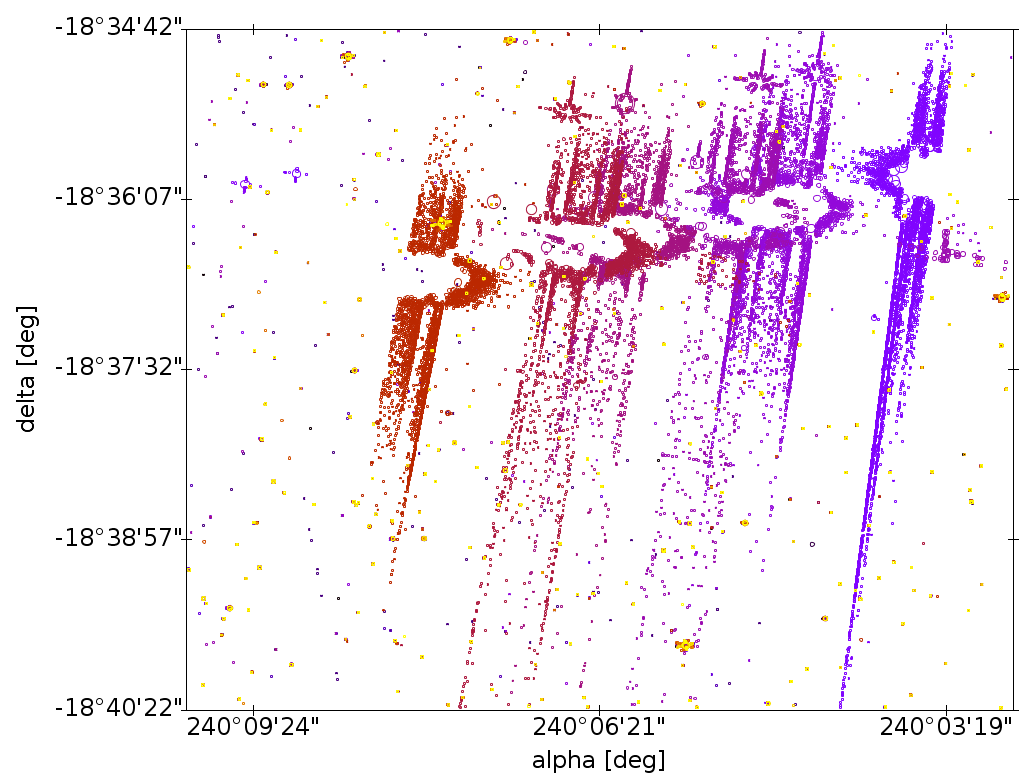}
\caption{Spurious detections from several consecutive Saturn transits. 
%projected onto the celestial sphere using the initial attitude reconstruction OGA1 described in Sect.~\ref{sec:oga}. 
The plot shows more than 22\,000 
detections during 33 scans and how the planet transits pollute an extended sky region.
\label{fig:xmSsoSpuriousExamples}}
\end{figure}

\begin{figure}
\centering
\includegraphics[width=1.0\columnwidth]{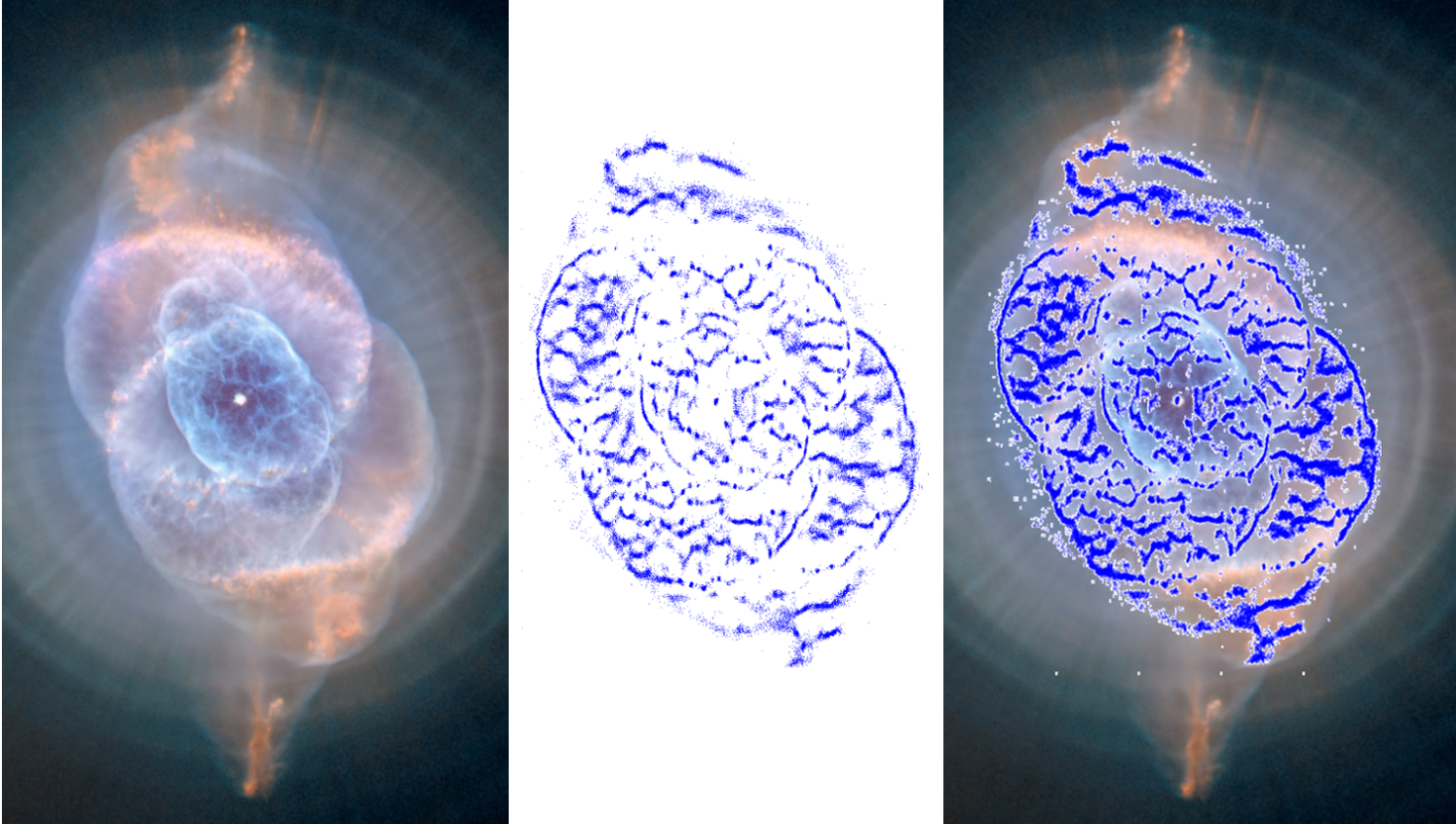}
\caption{Cat's Eye Planetary Nebula (NGC 6543) observed with the Hubble Space Telescope (left image) and 
as Gaia detections (the 84\,000 blue points in middle and right images) (Credit: Photo: NASA/ESA/HEIC/The Hubble 
Heritage Team/STScI/AURA).
\label{fig:xmNgcSpuriousExamples}}
\end{figure}

The dominating categories of spurious detections found in the data so far are:

\begin{itemize}
\item Spurious detections around and along the diffraction spikes of sources brighter than about 16~mag. For very bright stars there may
be hundreds or even thousands of spurious detections in a single transit, especially 
along the diffraction spikes in the AL direction, see Fig.~\ref{fig:xmBsSpuriousExamples} for an extreme example.

\item Spurious detections in one telescope originating from a very bright source in the other telescope, due
to unexpected light paths and reflections within the payload.

\item Spurious detections from major planets. These transits can pollute
large sky regions with thousands of spurious detections, see Fig.~\ref{fig:xmSsoSpuriousExamples}, but they can be
easily removed.

\item Detections from extended and diffuse objects. 
Fig.~\ref{fig:xmNgcSpuriousExamples} shows that Gaia is actually detecting not only stars but also
filamentary structures of high surface brightness.
These detections are not strictly spurious, but require a special treatment,
and are not processed for Gaia-DR1.

\item Duplicated detections produced from slightly asymmetric images where more than
one local maximum is detected. These produce redundant observations and must be identified during the cross-match.

\item Spurious detections due to cosmic rays. A few manage to get through the
onboard filters, but these are relatively harmless as they happen randomly
across the sky.

\item Spurious detections due to background noise or hot CCD columns. Most are caught onboard, so they are few and cause no serious trouble.
\end{itemize}

No countermeasures are yet in place for the last two categories, but this has
no impact on the published data, as these detections happen randomly on the sky
and there will be no corresponding stellar images in the astrometric (AF) CCDs.

%As commented above, the significant impact of the spurious detections is an issue recently identified. The current
%mitigation measures and software modules implemented are still under active development in both IDT and IDU. However, it
%is clear that, an effective mitigation scheme will be part of the cyclic data reduction with inputs from all downstream
%processes.

For Gaia-DR1 we identify spurious detections around bright source transits, either using actual Gaia detections of those or
the predicted transits obtained in the {scene}, and we select all the detections falling within a predefined set of
boxes centred on the bright transit. The selected detections are then analysed, and they are classified as
spurious if certain distance and magnitude criteria are met. These predefined boxes have been
parametrised with the features and patterns seen in the actual data according to the magnitude of the source.

For very bright sources (brighter than 6 mag) and for the major planets this model has been extended. For
these cases, larger areas around the predicted transits are considered and in both fields of view.

Identifying spurious detections around fainter sources (down to 16~mag) is more difficult, since there are often
only very few or none. In
these cases, a multiepoch treatment is required to know if a given detection is genuine or spurious -- i.e.\ checking if
more transits are in agreement and resolve to the same new source entry. These cases will be addressed in future data releases
as the data reduction cycles progress.

All in all, we have successfully identified the vast majority of spurious
detection, but some fraction (less than 20\%) still remains. These residual
detections will have entered the astrometric processing, but, again, the vast
majority will not produce a sensible solution and therefore not appear in
Gaia-DR1.

Finally, spurious new sources can also be introduced by excursions of the on-ground attitude reconstruction used for projecting the detections onto the sky (i.e.~short intervals of large errors in OGA1), leading to misplaced detections. 
Therefore, the attitude is carefully analysed to identify and clean up these excursions before the cross-match is run.

%% file: xm_coord.tex
\subsection{Sky coordinates determination\label{sec:xm:coord}}

The images detected on board, in the real-time analysis of the sky mapper data, are propagated to their expected
transit positions in the first strip of astrometric CCDs, AF1, i.e.\ their transit time and AC column are extrapolated and expressed as
a {reference acquisition pixel}. This pixel is the key to all further onboard operations and to the identification of the
transit. For consistency, the cross-match does not use any image analysis other than the onboard detection, and
is therefore based on the reference pixel of each detection, even if the actual image in AF1 may be slightly offset from it.
This decision was made because we in general do not have the same high-resolution SM and AF1 images on ground as were 
used on board.

\begin{figure}
\centering
\includegraphics[width=0.95\columnwidth]{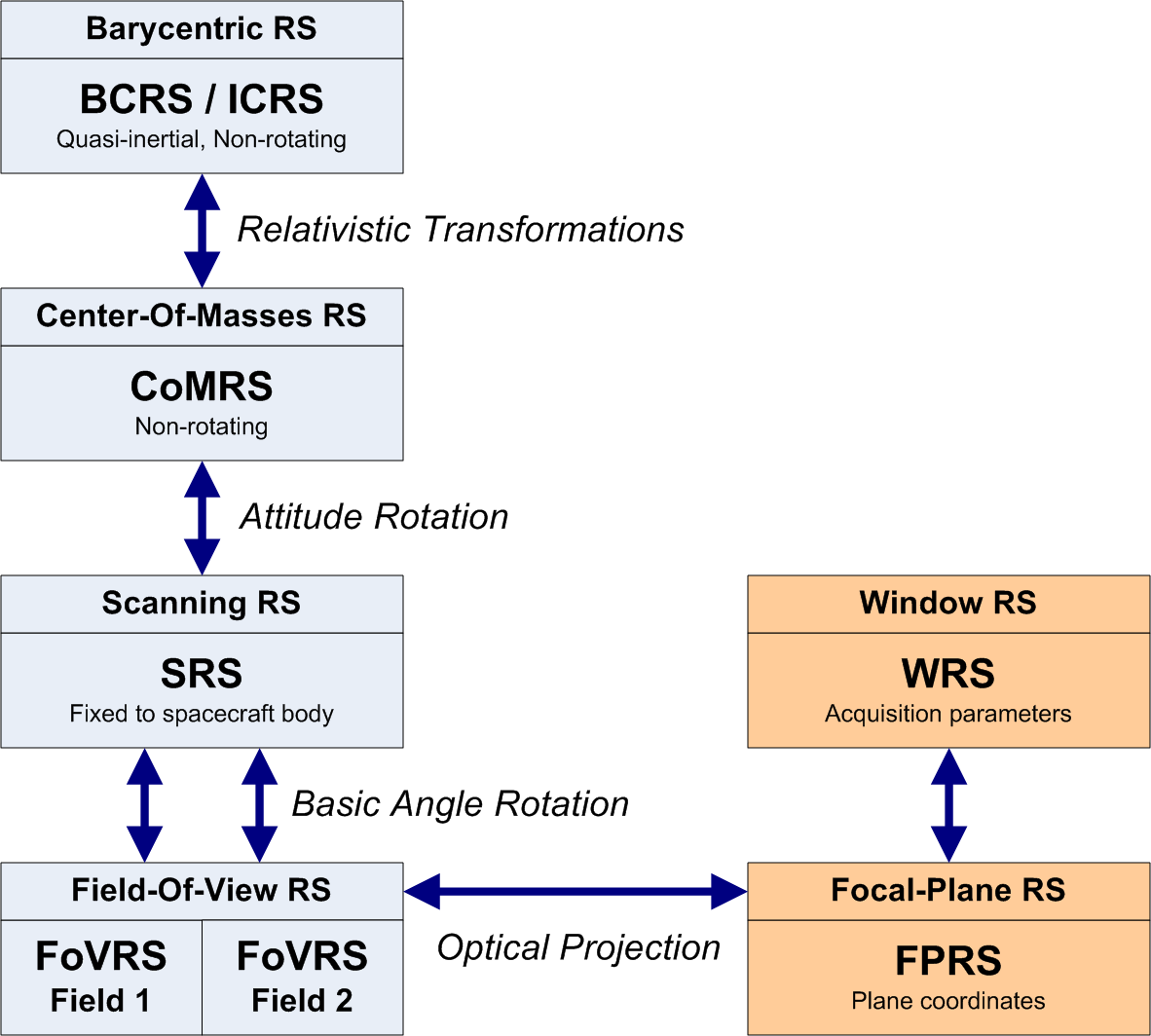}
\caption{Overview of the several reference systems used in pre-processing. From barycentric coordinates to 
the system used for the acquisition parameters of the observations within each CCD of the focal plane.
The transformations on the left are of a general, large scale nature, while the ones on the right involve the
detailed properties of the Gaia mirrors and focal plane.
\label{fig:xmReferenceSystems}}
\end{figure}

The first step of the cross-match is the determination of the sky coordinates of the Gaia detections,
but of course only for those considered genuine. 
As mentioned, the sky coordinates are computed using the reference acquisition pixel in AF1, and the precision is
therefore limited by the pixel resolution as well as by the precision of the onboard image parameter determination. 
The conversion from the observed positions on the focal plane to celestial coordinates, e.g.\ right 
ascension and declination, involves several steps and reference systems as shown in Fig.~\ref{fig:xmReferenceSystems}.

%SK:

The reference system for the source catalogue is the barycentric celestial
reference system (BCRS/ICRS), which is a quasi-inertial, relativistic reference
system non-rotating with respect to distant extra-galactic objects. Gaia
observations are more naturally expressed in the centre-of-masses reference
system (CoMRS) which is defined from the BCRS by special relativistic
coordinate transformations.  This system moves with the Gaia spacecraft and is
defined to be kinematically non-rotating with respect to the BCRS/ICRS.  BCRS
is used to define the positions of the sources and to model the light
propagation from the sources to Gaia. Observable proper directions towards the
sources as seen by Gaia are then defined in CoMRS. The computation of
observable directions requires several sorts of additional data like the Gaia
orbit, solar system ephemeris, etc.  As a next step, we introduce the scanning
reference system (SRS), which is co-moving and co-rotating with the body of the
Gaia spacecraft, and is used to define the satellite attitude.  Celestial
coordinates in SRS differ from those in CoMRS only by a spatial rotation given
by the attitude quaternions. The attitude used to derive the sky coordinates
for the cross-match is the initial attitude reconstruction OGA1 described in
Sect.~\ref{sec:oga}.

We now introduce separate reference systems for each telescope, called the
field of view reference systems (FoVRS) with their origins at the centre of
masses of the spacecraft and with the primary axis pointing to the optical
centre of each of the fields, while the third axis coincides with the one of
the SRS. Spherical coordinates in this reference system, the already mentioned
field angles
($\eta, \zeta$), are defined for convenience of the modelling of the
observations and instruments. Celestial coordinates in each of the FoVRS differ
from  those in  the SRS only  by  a fixed nominal spatial rotation around the
spacecraft rotation axis, namely by half the basic angle of 106\fdg5.

Finally, and through the optical projections of each instrument, we reach the focal plane reference system (FPRS),
which is the natural system for expressing the location of each CCD and each pixel. Figure~\ref{fig:fpa} indicates
(small yellow circles) the origin of the FoVRS field angles for each of the two telescopes as projected on the focal plane. 
It is also convenient to extend the FPRS to express the relevant parameters of each detection, specifically the field of view, CCD, gate,
and pixel. This is the window reference system (WRS). In practical applications, the relation between the WRS and the
FoVRS must be modelled. This is done through a {geometric calibration}, expressed as corrections to nominal
field angles as detailed in \citet[][Sect.~3.4]{2012A&A...538A..78L}. The geometric calibration used in the daily pipeline is derived by the First-Look system in the `one-day astrometric solution' (ODAS), see Sect.~\ref{sec:valid:fl}.

%% file: xm_mcg.tex
\subsection{Determination of Match Candidate Sources and Groups\label{sec:xm:mcg}}

Once we have the observation sky coordinates, we compare them with a list of sources extracted from the working
catalogue. These sources cover the region of the sky seen by Gaia in the relevant time interval, 
propagated to the epoch of observation.

Solar system objects are not included in the first Gaia data release. They are not identified during the pre-processing cross-match,
and will therefore instead lead to the creation of new sources. As each of these new sources will rarely be observed more
than once, they are automatically excluded from publication for this data release.

Candidate sources are selected based on a pure distance criterion, and using the same criterion for all sources. For Gaia-DR1, a match-radius of 1\farcs5 was
used, balancing the quality of IGSL with the wish to avoid too many
ambiguities.
The decision of only using the distance was taken
because the position of a source changes slowly and predictably, whereas other parameters as the magnitude may change in
an unpredictable way. 

The result of this first cross-match step is a set of the so-called match candidates, identifying all the possibly matching
sources from the working catalogue for each individual observation. Together with the match candidates, an auxiliary
table is also produced to track the links created for each source.

\begin{figure}
\centering
\includegraphics[width=0.85\columnwidth]{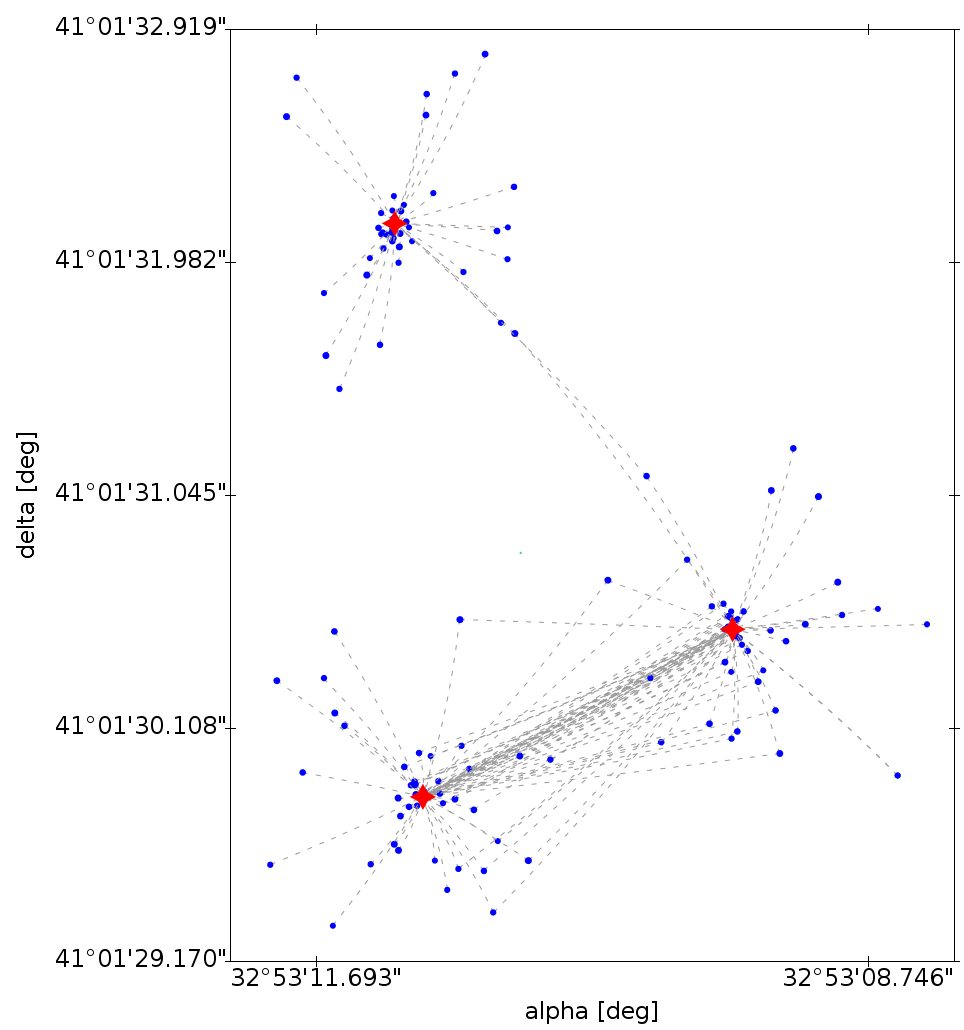}
\caption{Example of a match candidate group; in this case comprising three catalogue sources and about 180~observations. Blue dots correspond to observations, red 
dots are sources, and the dashed lines represent the match candidate source links.
\label{fig:xmMcgExample}}
\end{figure}

Next, the match candidates are first grouped to create self-contained match candidate groups of observations. These groups are
created applying a clustering algorithm over the detections according to their source candidates.
Figure~\ref{fig:xmMcgExample} shows an example. The objective is to determine isolated groups of detections, i.e.~groups not
related through common source candidates, located in a rather small and confined sky region, which can thus be  processed
independently from the other observations. This partitioning is essential in order to distribute the cross-match processing
in a computer cluster avoiding the usual issues of generic partitioning schemes, i.e.~problems in the treatment of detections close to the
region boundaries and in the handling of high proper motion stars which cannot be easily bound to any fixed, pre-defined region.

%% file: xm_solve.tex
\subsection{Match resolution\label{sec:xm:solve}}

The final step of the cross-match is the most complex, resolving the final matches and 
consolidating the final new sources. We distinguish three main cases to solve:

\begin{itemize}
\item Duplicate matches: when two (or more) detections close in time 
are matched to the same source. This will typically be either newly resolved
binaries or spurious double detections.

\item Duplicate sources: when a pair of sources from the catalogue have never been observed simultaneously, thus never 
identifying two detections within the same time frame, but having the same matches. 
This can be caused by double entries in the working catalogue.

\item Unmatched observations: observations without any valid source candidate.
\end{itemize}

For Gaia-DR1, the resolution algorithm has been based on a nearest-neighbour solution where the conflict between two given observations is resolved independently from the other observations 
included in the group. The main disadvantage of this approach is that the number of new sources created, when more 
than two observations close in time have the same source as primary match, is not minimised.

As a result of a number of issues with the IGSL and with residual spurious detections,
many stars with high proper motion, especially above 3\arcsec/yr, were not
correctly resolved and are therefore missing in Gaia-DR1.

Forthcoming Gaia data releases will be based on more sophisticated clustering solutions where all the 
inter-relations of the observations contained in each group will be taken into account to generate the best resolution.

%% file: bam.tex
\section{Basic-angle monitor (BAM)\label{sec:bam}}

The Gaia measurement principle is that differences in the transit time between stars observed by each telescope can be translated into angular measurements. All these measurements are affected if the basic angle (the angle between telescopes, $\Gamma = 106\fdg5$) is variable. Either it needs to be stable, or its variations be known to the mission accuracy level ($\approx$1~$\mu$as).

Gaia is largely self-calibrating (calibration parameters are estimated from
observations). Low frequency variations ($f < 1 / 2P_{\rm rot}$) can
be fully eliminated by self-calibration. High frequency random variations are
also not a concern because they are averaged during all transits. However,
intermediate-frequency variations are difficult to  eliminate by
self-calibration, especially if they are synchronised with the spacecraft spin
phase, and the residuals can introduce systematic errors in the astrometric
results \citep[][Sect.~2]{2016A&A...586A..26M}. Thus, such intermediate-frequency changes need to be monitored by
metrology.

\begin{figure}
  \includegraphics[width=\hsize]{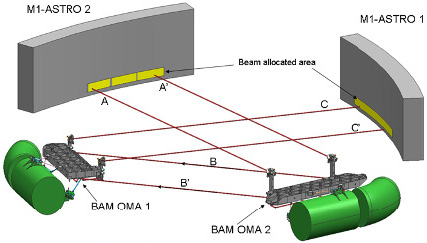}
\caption{BAM overview. The BAM is a laser interferometer that injects two beams in each telescope entrance pupil. In this way, an interference pattern is produced for each telescope in the common focal plane. The relative shift of the patterns at the CCD level is related to changes in the basic angle between the telescopes. Credit: Airbus Defence \& Space.
\label{fig:gaiaWorkingPrinciple}}
\end{figure}

The BAM device is continuously measuring differential
changes in the basic angle. It basically generates one artificial fixed star
per telescope by introducing two collimated laser beams into each primary
mirror (see Fig.~\ref{fig:gaiaWorkingPrinciple}). The BAM is composed of two
optical benches in charge of producing the interference pattern for each
telescope. A number of optical fibres, polarisers, beam splitters and mirrors
are used to generate all four beams from one common light source. See
\citet{2012SPIE.8442E..1RG} for further details. Each Gaia telescope then
generates an image on the same dedicated BAM CCD (Fig.~\ref{fig:fpa}), which is an interference pattern due to the coherent input
light source. The relative AL displacement between the two fringe
patterns is a direct measurement of the basic-angle variations.

\subsection{Data collection}

BAM data acquisition is driven by the time needed to obtain
high-signal-to-noise fringes without saturation. It is independent of the
spacecraft attitude (spinning or not, scanning law etc.) and control mode
(micropropulsion or chemical thrusters). In this way, BAM
frames are always acquired, provided the laser is on and the CCD readout is in
a normal operational mode. The acquisition period is fixed, and has been kept
to 23.5 seconds.
Major events, such as safe modes, payload data handling outages, and payload
decontamination campaigns have occasionally interrupted the BAM data flow for a
while. Routine calibration activities and ground station issues sometimes
introduce additional small gaps.

The interference patterns cover a small part of the CCD, so some windowing is needed. For each field of view, a pattern of 1000 AL $\times$ 80 AC samples is acquired and downlinked in two telemetry packets. Each of these samples is asymmetrical, the AC size being 12 times larger than AL. This scheme is applied because the fringes are well aligned to the AC direction. The physical area of the windows on the detector is almost square (10 mm $\times$ 9.6 mm).

\subsection{BAM data model and fit}

Several strategies have been proposed to analyse the BAM data. Amongst them are cross-correlation, Fourier transform and forward modelling. The latter one has been selected for the BAM analysis in the pre-processing (specifically within IDT). It uses a mathematical model to represent the BAM image, which is then fitted using a least-squares algorithm. The noise model uses two components: Poissonian shot noise and CCD read-out noise. Maximum likelihood performance could in principle be achieved if the model provided a faithful representation of the real patterns. This method is much slower than the others, but provides greater flexibility to accommodate future improvements. In practice, a balance is required to provide high-quality results in a reasonable time and to avoid overfitting by a too complex function.

The mathematical model and fitting process is described in detail in Mora et al. (2016, in preparation) and \citet{2014SPIE.9143E..0XM,2014EAS....67...65M}. It basically combines the light of two perfectly Gaussian laser beams using plane-parallel equally spaced aberration-free fringes:

\begin{equation}
  I = I_{G1} + I_{G2} + 2\sqrt{I_{G1} I_{G2}} \cos \delta
%\label{eq:}
\end{equation}

\noindent where $I$ is the irradiance at a given point on the detector, $I_{G1}$ and $I_{G2}$ the individual beam irradiances, the third term provides the interference and $\delta$ is the optical path difference

\begin{equation}
  \delta = \frac{2 \pi B d}{\lambda f}
%\label{eq:}
\end{equation}

\noindent where $B$ is the interferometer baseline, $\lambda$ the wavelength, $f$ the focal length and $d$ the distance to the reference fringe. This basic model is somewhat extended to integrate the irradiance over each pixel, provide an analytical representation of the derivatives, and account for the read-out smearing background (there is no shutter).

\begin{figure}
  \includegraphics[width=\hsize]{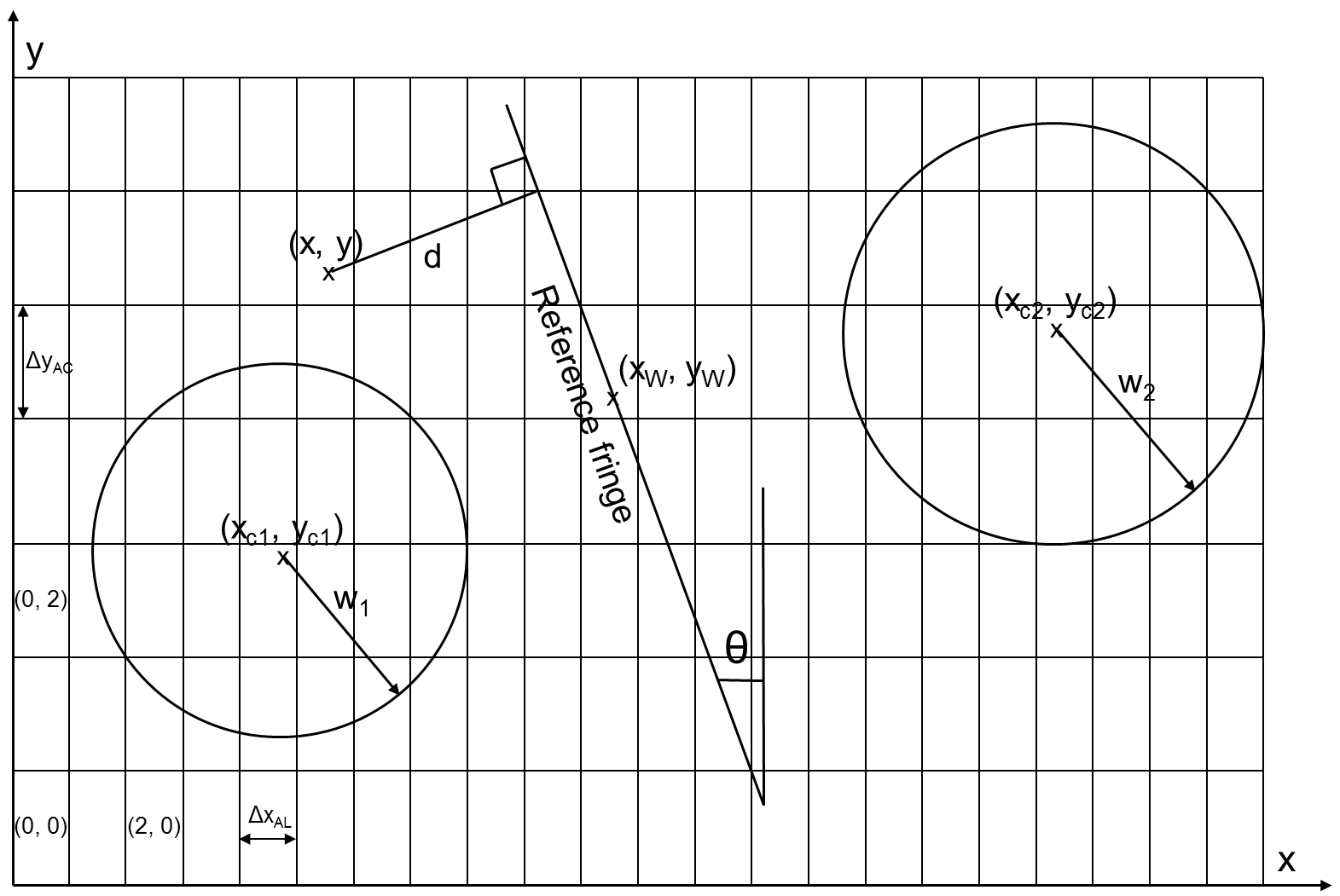}\\
  \vspace{-0.3cm}\\
  \includegraphics[width=\hsize]{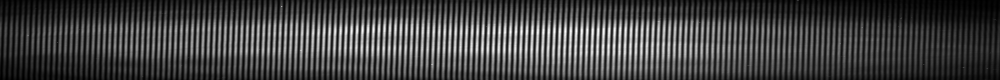}\\
  \vspace{-0.3cm}\\
  \includegraphics[width=\hsize]{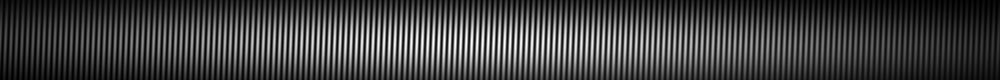}\\
  \vspace{-0.1cm}\\
  \includegraphics[width=\hsize]{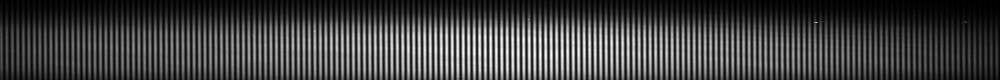}
  \vspace{-0.3cm}\\
  \includegraphics[width=\hsize]{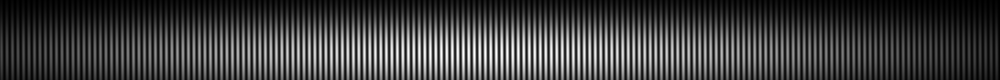}
\caption{Top: BAM image model. The perfect plane-parallel fringes produced by the interference of two Gaussian beams are integrated for each pixel. The most important parameter is the AL position (phase) of the reference fringe. Middle: Real vs.\ model BAM patterns for the preceding field of view. Bottom: Real vs.\ model BAM patterns for the following field of view.
\label{fig:BamModelVsReality}}
\end{figure}

The total number of free parameters per fringe pattern is 11:~two Gaussian beam locations and peak fluxes, a common waist (width), additive ``sky'' constant, reference fringe AL location (phase), rotation angle, and fringe period. The reference fringe phase is the key output variable, the others just being nuisance parameters, useful to monitor the variability of the fringe patterns. The fringe period is $\simeq$5.5~pix, ensuring appropriate sampling above the Nyquist limit. Typical centroiding errors are $\simeq$98~$\mu$pix and $\simeq$52~$\mu$pix for single preceding and following field of view patterns, respectively. We note that the optical path for the preceding field of view contains many more mirrors, hence the lower photon counts. This precision is sufficient for DR1 purposes. For subsequent data releases, limited tests suggest that the precision could be improved up to a factor $\simeq$2$\times$, reaching the Cram\'er-Rao ultimate limit, if the cosmic rays in the input images are appropriately filtered. Figure~\ref{fig:BamModelVsReality} shows the model design in a graphical way, together with some examples of observed and model interferograms.

\subsection{BAM daily data processing}

There are three systems forming 
the BAM daily pipeline running at ESAC: MIT, IDT and First Look.

As mentioned in Sect.~\ref{sec:daily}, MIT reconstructs the telemetry stream, 
while the data processing is done by IDT.

IDT assembles the different BAM telemetry packets and generates a high-level object with the fit results, which basically contains the flux and all fit parameters and their formal errors. An output object is always produced, even when no convergence has been achieved, which is indicated by processing flags in the object. The process is fully automatic, the only manual operation is the occasional update of auxiliary calibration tables.

\begin{figure}
\begin{center}
\includegraphics[trim={2.5cm 3.5cm 2cm 3.0cm},clip,width=\hsize]{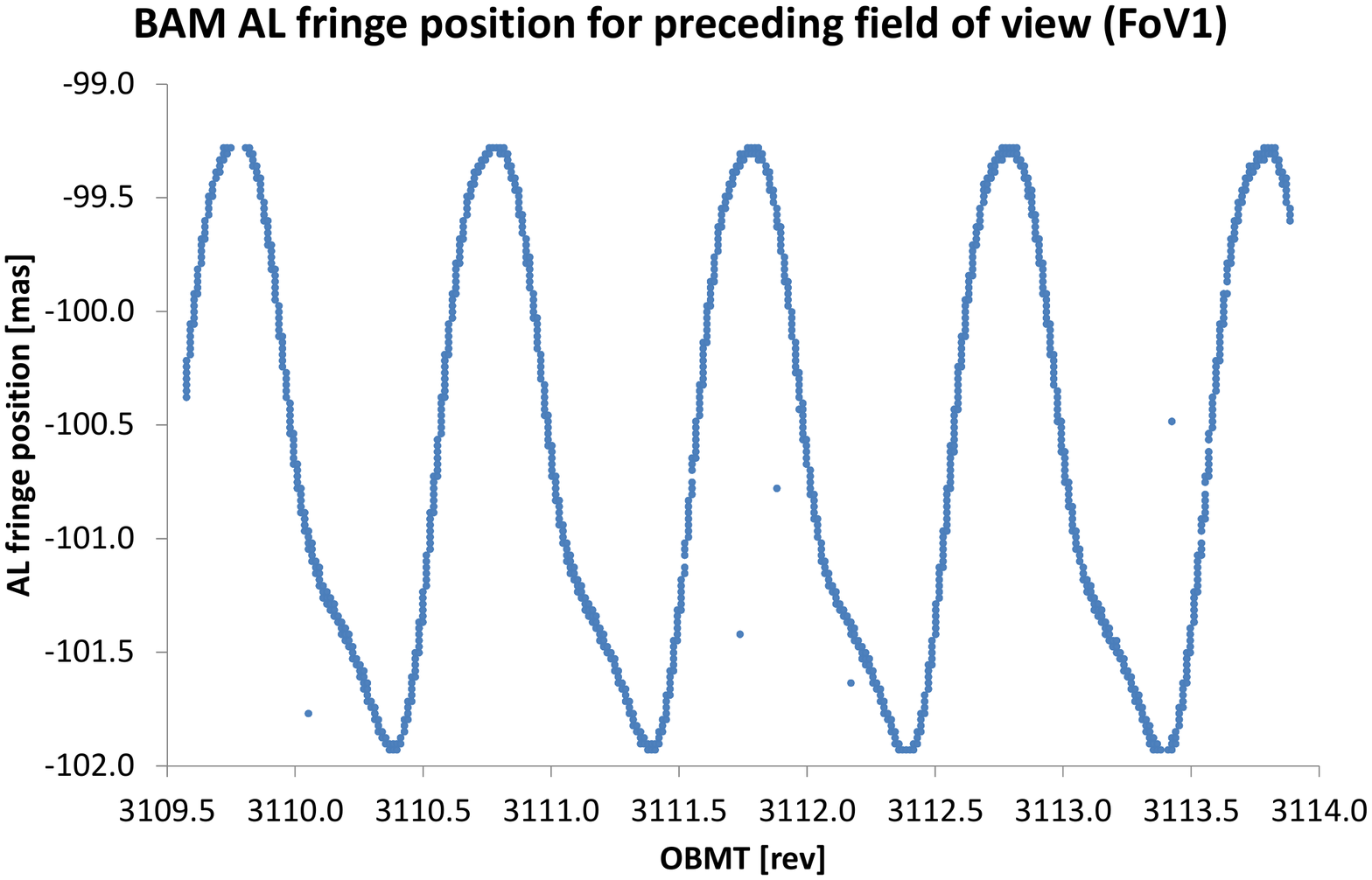}
\includegraphics[trim={2.5cm 3.5cm 2cm 3.0cm},clip,width=\hsize]{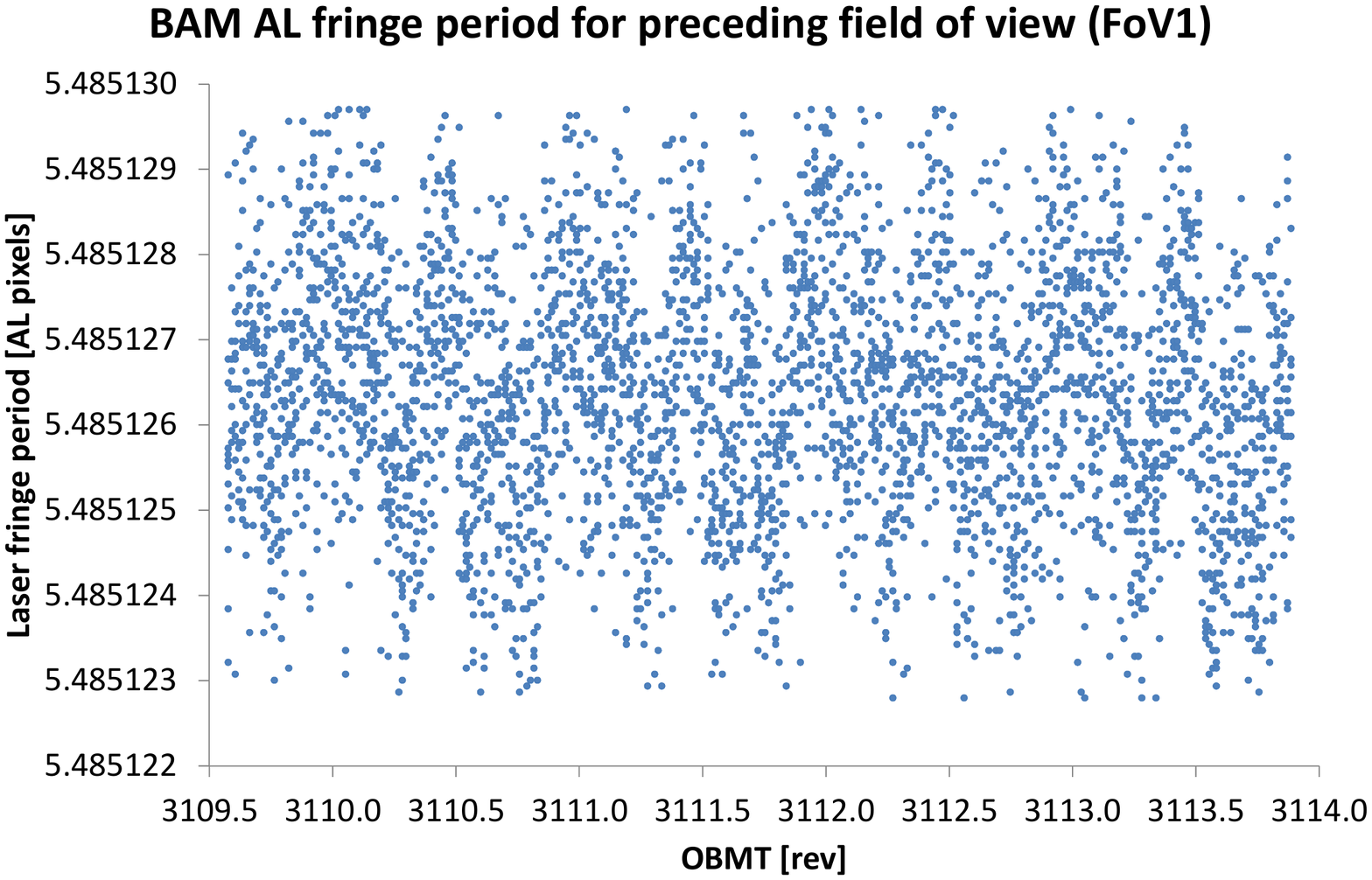}
%\includegraphics[width=\hsize]{figures/sfunc740_alFringe_Fov1}
%\includegraphics[width=\hsize]{figures/sfunc740_laserWavelength_Fov1}
%\includegraphics[width=\hsize]{figures/sfunc7020_absolute}
%\includegraphics[width=\hsize]{figures/sfunc740_alFringe_Fov2}
%\includegraphics[width=\hsize]{figures/sfunc740_laserWavelength_Fov2}
% Data taken from DFLE report
% FL Solution ID: 3247974940636348431
% Report time-range: OBMT: 3109.56913 – 3113.89413 [revs]
% UTC: 2015-12-15T05:38:14 – 2015-12-16T07:35:14
% FL version: 19.1.0
% FL jar file: FL-19.1.0.jar
% IDT Run Number: 752
% Generated on Thu Dec 17 15:20:10 UTC 2015
\end{center}
\caption{BAM daily First-Look analysis. The BAM data quality is monitored on a daily basis using several metrics implemented in the First-Look system. The top panel shows the evolution of the fringe phase for the preceding field of view during four spacecraft revolutions beginning on 2015-12-15. The significant ($\approx$1\,mas) amplitude of the periodic variations is apparent. The bottom panel shows the evolution of a secondary (nuisance) parameter, namely the average fringe period, during the same interval ($\approx$1\,ppm variability in this case). The plots include average values computed by First Look but have been adapted for better visualisation.
\label{fig:bamFirstLook}}
\end{figure}

First Look provides multiple diagnostics using BAM data for quick analysis and
payload health assessment. They primarily consist of histograms and time
evolution plots for each fit parameter. Long-term trend analysis plots are also
created. The (single) daily basic-angle value estimated by the ODAS (see
Sect.~\ref{sec:valid:fl}) is also provided for comparison.
Figure~\ref{fig:bamFirstLook} shows an example of typical daily evolution plots
of the fringe phase and fringe period for the preceding field of view.

First Look is the end of the BAM daily processing. The results are typically analysed by humans, but not used by further downstream automatic systems. However, the IDT output has been analysed in much detail (outlier rejection, discontinuity correction, trend removal and Fourier analysis) and the results have been introduced as an integral part of the astrometric processing, see Mora et al. (2016, in preparation). Many features have been identified in the BAM data, most notably the significant $\approx$1~mas amplitude heliotropic periodic oscillations. They are discussed in detail in Mora et al. (2016, in preparation). This is a factor $\approx$250 above the pre-launch expected basic-angle stability. However, the repeatability of the major periodic features, the exceptional BAM measurement precision ($\simeq$0.5$\mu$as per 10-20~min interval) and a careful signal modelling (see Lindegren et al.\ 2016, this volume) ensure the basic-angle variations are not driving the systematic errors budget for this data release.

%% file: valid_mon.tex
\section{Validation\label{sec:valid}}

All systems presented here must be carefully monitored during their
execution and their outputs validated in detail. 

Four independent software systems are in use for quick monitoring and
validation of the spacecraft and the daily pipeline. They are briefly
described in the following four subsections. The first two are run within the
main daily pipeline at the Gaia Science Operations Centre at ESAC. The latter
two are run at the Turin data processing centre.

\subsection{Daily monitoring within IDT\label{sec:valid:mon}}

We have developed a near-real-time monitoring tool for the IDT system, with
both a web-based interface and automatic PDF reports with the main diagnostics.
The tool shows the overall progress of the processing, including the mission
time, the number of measurements and output records, the
processing performance, and a variety of other technical monitors. Besides, a
large number of plots and tables are generated. They illustrate statistics and
checks on the several data types, as well as the distribution
and correlations of several scientific quantities.  This monitoring is done
for the data covering roughly 24 hours, so that we can also see
possible trends or degradations in some processing -- either caused by
onboard or on-ground issues.

\begin{figure}
  \begin{center}
    \includegraphics[width=\columnwidth]{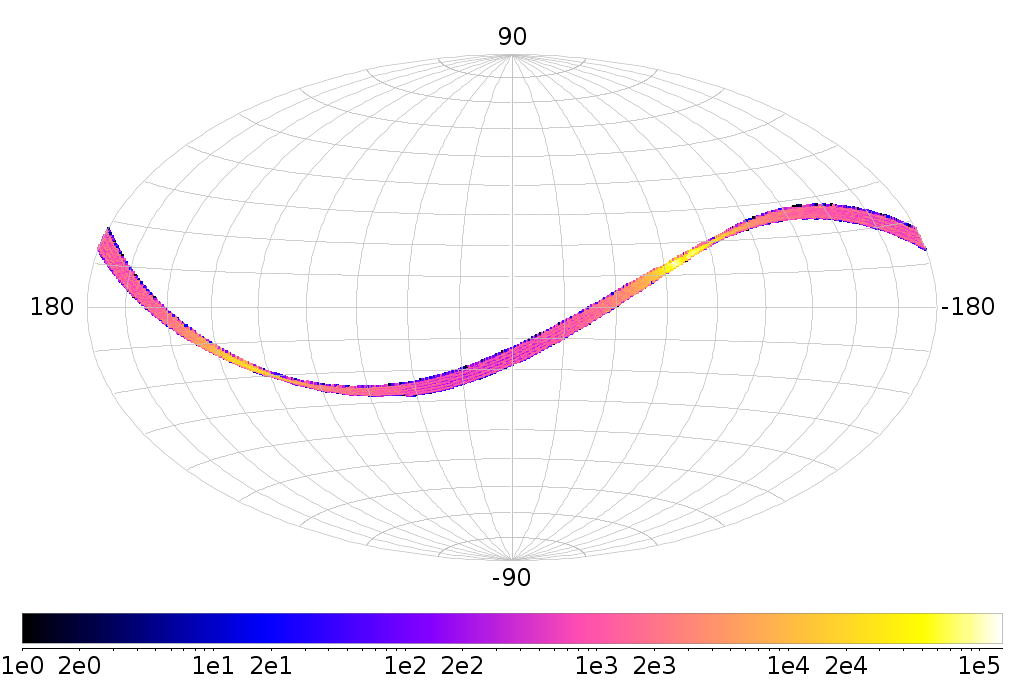}
    \caption{\label{fig:webmon_skymap}
      Sky map showing the region observed during about one day in equatorial
      coordinates. The colour scale indicates the density of transits
      per square degree, reaching 0.7~million in this example. This density
      is a combination of the actual star density and the satellite scanning law.
    }
  \end{center}
\end{figure}

\begin{figure}
  \begin{center}
    \includegraphics[width=\columnwidth]{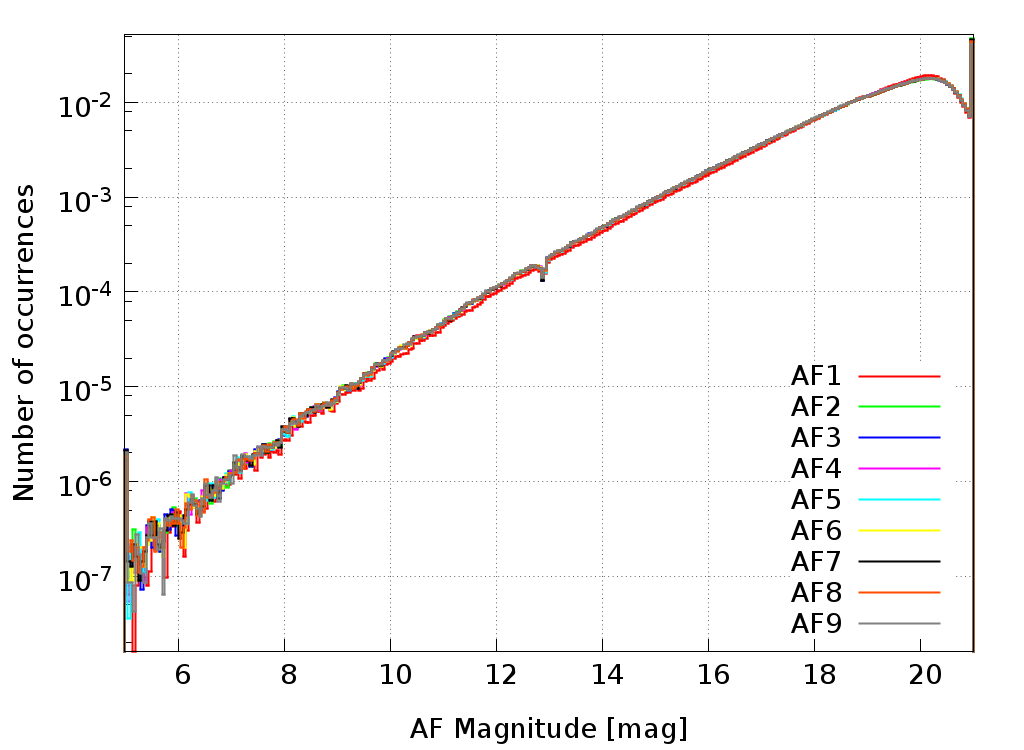}
    \caption{\label{fig:webmon_afg}
      Histogram of the fraction of transits per magnitude bin of 0.05~mag.
      The G magnitude was determined from IDT image parameters.
      Observations outside the range are included in the extreme bins.
    }
  \end{center}
\end{figure}

\begin{figure}
  \begin{center}
    \includegraphics[width=\columnwidth]{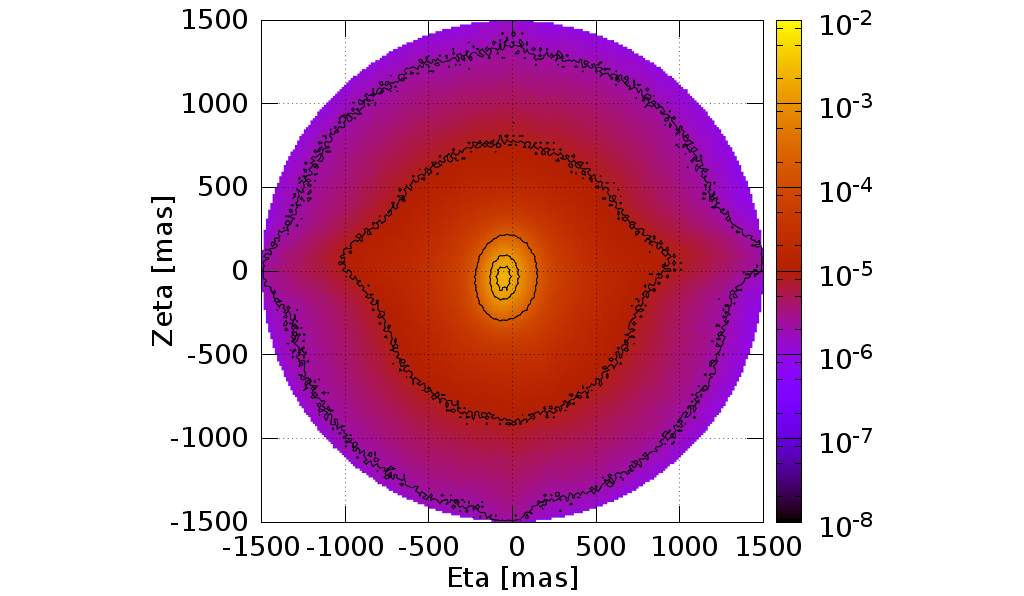}
    \caption{\label{fig:webmon_xmdist}
      Two-dimensional histogram showing the distribution of the match distances
      between detected positions and their corresponding catalogue sources,
      for the AL and AC field angles ($\eta$ and $\zeta$, respectively).
      Contour lines are also shown, which help detecting features in the data
      or issues in the instrument or in the data processing.
    }
  \end{center}
\end{figure}

As an example of the mentioned diagnostics, Fig.~\ref{fig:webmon_skymap} shows
the sky region observed (and processed) during about one day, indicating the
density of measurements (transits) per square degree.  Since Gaia spins four
times a day and has two telescopes, the density of transits can reach eight
times the actual star density in the sky.  As another example,
Fig.~\ref{fig:webmon_afg} shows the distribution of the number of transits per
magnitude bin for the nine astrometric CCD strips.  The small dip at 13~mag is
due to the transition between two-dimensional windows with PSF fitting for the brighter
stars, and the LSF fitting of one-dimensional windows for the fainter sources. 
The latter is displaced slightly to the right, since the difference in PSF and LSF normalisations makes LSF based
fluxes systematically fainter than PSF based ones. These differences are of course corrected in the photometric calibrations.
Finally, Fig.~\ref{fig:webmon_xmdist} shows the distance between detections and
catalogue sources for the preliminary cross-match, which is obviously limited
to the match radius used (currently set to 1\farcs 5).

Overall, the daily IDT monitoring system provides a large number of technical and scientific diagnostics
and monitors. Besides the examples shown, it includes more plots with photometric distributions
for different bands, the attitude correction done and the rates obtained, the variations and fluxes from the
basic angle monitoring system, electronic bias and readout noise, astrophysical background, centroids obtained
in the main image parameter determination (as well as goodness-of-fit and formal errors), and figures
from the preliminary cross-match such as the number of transits not found in the catalogue (leading
to new entries). Most of these diagnostics are separately available for each CCD, which allows determining
problems in the instrument or in the on-ground calibrations.

%% file: valid_fl.tex
\subsection{First Look: Extended daily instrument health checks and data validation\label{sec:valid:fl}}

Astrometric space missions like Gaia have to simultaneously determine a tremendous number of parameters concerning astro\-metry and other stellar properties, the attitude of the satellite as well as the geometric, photometric, and spectroscopic calibrations of the instrument. 

To reach the inherent precision level of Gaia, many months of observational data
have to be incorporated in a global, coherent, and interleaved data reduction.
Neither the instrument nor the data health can be verified at the
desired level of precision by standard procedures applied to typical space
missions, e.g.~by the IDT monitoring described in the previous subsection.
Obviously it is undesirable not to know the measurement precision and
instrument stability until more than half a year of the mission has elapsed. If
any unperceived, subtle effect would arise during that time this would affect
all data and could result in a loss of many months of data.

For this reason a rapid {First Look} was installed to daily judge the data at a more sophisticated level. It aims at a quick discovery of delicate changes in the spacecraft and payload performance. Its main goal is to prompt onboard actions if scientific need arises, but it also aims at identifying oddities and proposing potential improvements in the on-ground data reduction. For these purposes, tens of thousands of higher-level diagnostic quantities are derived daily from:
\begin{itemize}
	\item The satellite housekeeping data;
    \item Auxiliary data such as onboard status data which are needed to allow a proper science data reduction on ground, and onboard processing counters which allow  us to check the sanity of the onboard processing;
    \item Photometric and spectroscopic science data;
    \item BAM data as described in Sect.~\ref{sec:bam};
    \item Astrometric science data;
    \item The variety of IDT products derived from all these data.
\end{itemize}    

\begin{figure}
  \begin{center}
    \includegraphics[width=\columnwidth]{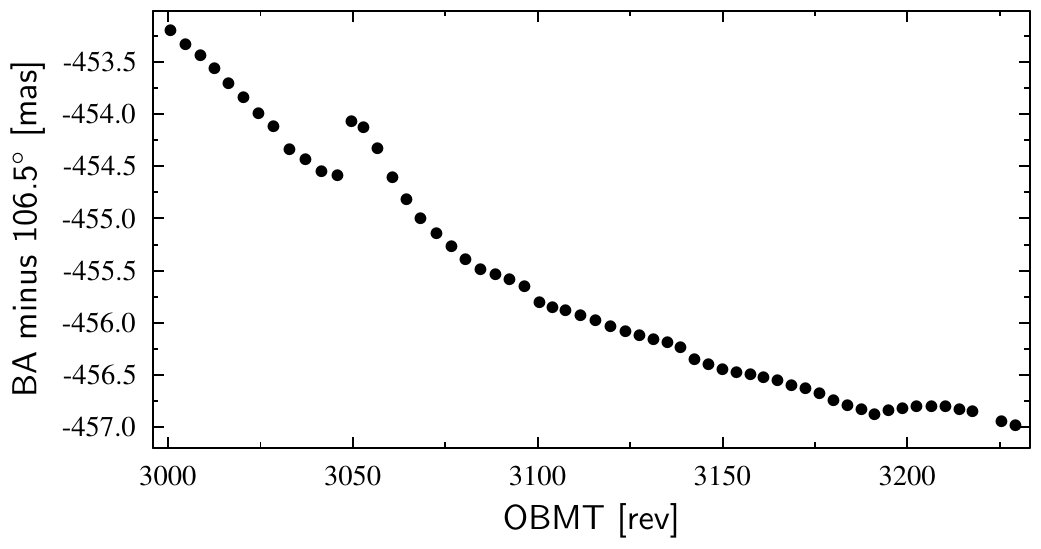}
    \caption{\label{fig:ODAS-basic-angle}
      Long-term trend of the ODAS-derived basic angle (BA) over two months of mission time. The ODAS determines the mean angular location of each CCD in each field of view with micro-arcsecond precision every day (the basic-angle oscillations shown in e.g.~Fig.~\ref{fig:bamFirstLook} are averaged out in the process), from which an absolute basic angle can be derived (the BAM performs only relative angular measurements). The plot reveals a harmless general gradient of the order of 1--2~mas per month, which corresponds to a relative movement of the edges of the main mirrors of the two telescopes by just a few nanometers per month. This trend is presumably due to seasonal temperature variations (caused by the eccentricity of the Earth orbit; the levelling-off of the curve at right coincides with the approach to the perihelion) and to ageing of materials in the space environment. The obvious jump and subsequent relaxation is caused by a thermal disturbance due to an unplanned 17-hour switch-off of the onboard payload data handling unit on November 29, 2015. Some of the smaller wiggles in the curve can also be traced to variable onboard conditions, such as e.g.~a varying star density on the sky scanned by Gaia.   
    }
  \end{center}
\end{figure}

The regular products and activities of the First-Look system and team include:
\begin{itemize}
	\item Most of the detailed daily calibrations described in Sect.~\ref{sec:ipd}; they are mainly produced in an automated fashion.
	\item The ODAS which allows us to derive a high-precision on-ground attitude, high-precision star positions, and a very detailed daily geometric calibration of the astrometric instrument. The ODAS is by far the most complex part of the First-Look system.
	\item Astrometric residuals of the individual ODAS measurements, required to assess the quality of the measurements and of the daily astrometric solution. 
	\item An automatically generated daily report of typically more than 3000 pages, containing thousands of histograms, time evolution plots, number statistics, calibration parameters etc.; a single example is given in Fig.~\ref{fig:ODAS-basic-angle}.
	\item A daily manual assessment of this report. This is made possible in about 1--2 hours by an intelligent hierarchical structure, extended internal cross-referencing and automatic signalling of apparently deviant aspects.
	\item Condensed weekly reports on all findings of potential problems and oddities. These are compiled manually by the so-called `First Look scientists team and the wider `payload experts group' (about two dozen people). If needed, these groups also prompt actions to improve the performance of Gaia and of the data processing on ground. Such actions may include telescope refocusing, change of onboard calibrations and configuration tables, decontamination campaigns, improvements of the IDT configuration parameters, and many others. 
	\item Manual qualification of all First-Look data products used in downstream data processing (attitude, source parameters, geometric instrument calibration parameters). 
\end{itemize}
	
In this way the First Look ensures that Gaia achieves the targeted data quality, and also supports the cyclic processing systems by providing calibration data. In particular the daily attitude and star positions are used for the wavelength calibration of the photometric and spectroscopic instruments of Gaia. In addition, the manual qualification of First-Look products helps both IDT and the cyclic processing teams to identify and discriminate healthy and (partly) corrupt data ranges and calibrations. This way, poor data ranges can either be omitted or subjected to special treatment (incl.~possibly a complete reprocessing) at an early stage. 

%% file: valid_avubam.tex
\subsection{AVU/BAM \label{sec:valid:bamavu}}

The BAM module \citep{2014RMxAC..45...35R} of the `astrometric verification unit' (AVU) is a completely independent counterpart of the BAM analysis by IDT/First Look described in Sect.~\ref{sec:bam}. Its purpose is to monitor on a daily basis the BAM instrument and the basic-angle variation independently from IDT, to provide periodic and trend analysis on short and long timescales, and to finally provide calibrated measurements of the basic-angle variations, as well as a model of their temporal behaviour. 
Here we only refer to the daily monitoring provided %by AVU/BAM 
in the context of validation of the IDT results.

In addition to producing time series of the fringe phase variations, AVU/BAM also makes measurements of other basic quantities, characteristic of the BAM instrument, like fringe period, fringe flux, and fringe contrast. Their temporal variations are also monitored and analysed to support the interpretation of the
basic-angle variations.

The AVU/BAM pipeline performs two different kinds of analyses: the first is based on daily runs, the second is focused on overall statistics 
on a weekly/monthly basis. The pipeline produces automatic daily reports.

Three different methods are utilised for computing basic-angle variations: one similar to the IDT approach, and one- and two-dimensional versions, respectively, of a direct measurement.
A description of the AVU/BAM system, including the three algorithms, will be given in Riva et al.~(2016, in preparation). The four independent results (three AVU/BAM plus IDT) agree quite well in the general character and shape of the 6-hour basic-angle oscillations, while the derived amplitudes differ at the level of 5~percent. It is as yet unknown which of the four methods (and of the implied detailed signal models fitted to the BAM fringe patterns) gives the most faithful representation of the relevant variations in the basic angle of the astrometric instrument.

%% file: valid_avuaim.tex
\subsection{AIM \label{sec:valid:aimavu}}

The `astrometric instrument model' (AIM) is a scaled-down counterpart of IDT and
First Look restricted to some astrometric elements of the daily processing; its
focus is on the independent verification of selected AF monitoring and
diagnostics, of the image parameter determination, and instrument modelling
and calibration. This separate processing chain is described in
\citet{2014SPIE.9150E..0KB}. 
In particular AIM provides image parameters through its own image
parameter estimation code, allowing routine comparisons with 
-- and thus external verification of -- 
IDT image location values and corresponding formal errors.
More details will be given in Busonero et al.~(2016, in preparation).

As stated in Sect.~\ref{sec:lsfpsf}, the reconstruction of the LSF and PSF are two of the Gaia key calibrations. 
For that reason, AIM implements its own independent PSF/LSF image profile models based in a one-dimensional case on 
a set of monochromatic basis functions, where the zero-order base is the sinc function squared.
The complete model includes the contribution of finite pixel size, modulation transfer function and CCD operation in TDI mode. 
The higher-order functions are generated by suitable combinations of the parent function and its derivatives according to a construction rule ensuring 
orthonormality. The spatially variable LSF/PSF is reconstructed as the sum of spatially invariant functions, with coefficients varying over the 
fields of view.
The polychromatic functions are built according to linear superposition of the monochromatic counterparts, weighted by the normalised detected 
source spectrum.
The model is briefly described in \citet{2013PASP..125..444G}; details will be given in Busonero et al.~(2016, in preparation).

%% file: appendix.tex
\begin{appendix}
\section{List of acronyms}
Below, we give a list of acronyms used in this paper.\hfill\\
\begin{tabular}{ll}\hline\hline 
\textbf{Acronym} & \textbf{Description} \\\hline
AC&Across Scan (direction) \\
ADC&Analogue-to-Digital Converter \\
ADU&Analogue-to-Digital Unit \\
AF&Astrometric Field (CCDs)\\
AIM&Astrometric Instrument Model \\
AL&Along Scan (direction) \\
AVU & Astrometric Verification Unit \\
BAM&Basic-Angle Monitor (Device) \\
BCRS&Barycentric Celestial Reference System \\
BP&Blue Photometer \\
CCD&Charge-Coupled Device \\
CDM&Charge Distortion Model \\
CoMRS&Centre of Mass Reference System \\
CTI&Charge Transfer Inefficiency \\
DPAC&Data Processing and Analysis Consortium \\
ESA&European Space Agency \\
ESAC&European Space Astronomy Centre  \\
FoVRS&Field of View Reference System \\
FPRS&Focal Plane Reference System \\
FWHM&Full Width at Half Maximum \\
Gaia-DR1&Gaia Data Release 1\\
GEPC&Gaia Ecliptic-Poles Catalogue  \\
HEALPix&Hierarchical Equal-Area iso-Latitude Pixelisation \\
ICRS&International Celestial Reference System \\
IGSL&Initial Gaia Source List \\
IDT&Initial Data Treatment  \\
LSF&Line Spread Function \\
MIT&Mission operations centre Interface Task \\
OBMT&On-Board Mission Timeline \\
ODAS&One-Day Astrometric Solution \\
OGA1&First On-Ground Attitude determination (IDT) \\
OGA2&Second On-Ground Attitude determination (ODAS) \\
PCA&Principal Component Analysis \\
PSF&Point Spread Function \\
RP&Red Photometer \\
RVS&Radial Velocity Spectrometer \\
SDSS&Sloan Digital Sky Survey \\
SM&Sky Mapper (CCDs)\\
SRS&Scanning Reference System\\
TCB&Barycentric Coordinate Time \\
TDI&Time-Delayed Integration (CCD) \\
WFS&WaveFront Sensor \\
WRS&Window Reference System \\\hline
\end{tabular} 
\end{appendix}